\documentclass[fleqn,10pt]{wlscirep}
\usepackage[utf8]{inputenc}
\usepackage[T1]{fontenc}
\usepackage{xcolor}
\usepackage{caption}
\usepackage{subcaption}
\usepackage[nice]{nicefrac}
\usepackage{soul}

\usepackage{amsmath}
\newcommand{\bs}[1]{\mathbf{#1}}

\newcommand{\blue}[1]{{\color{blue}#1}}
\newcommand{\ms}{\,\mathrm{m/s}}

\title{Dense Crowd Dynamics and Pedestrian Trajectories: A Multiscale Field Study at the Fête des Lumières in Lyon}

\author[1,*]{Oscar Dufour}
\author[2]{Huu-Tu Dang}
\author[3,5]{Jakob Cordes}
\author[4]{Raphael Korbmacher}
\author[3,*]{Mohcine Chraibi}
\author[2,*]{Benoit Gaudou}
\author[1,*]{Alexandre Nicolas}
\author[4,*]{Antoine Tordeux}
\affil[1]{Universite Claude Bernard Lyon 1, CNRS, Institut Lumière Matière, UMR5306, F-69100, Villeurbanne, France}
\affil[2]{UMR 5505 IRIT, Université Toulouse Capitole, Toulouse, France}
\affil[3]{Institute of Advanced Simulation, Forschungszentrum Jülich GmbH, Jülich, Germany}
\affil[4]{Fakultät für Maschinenbau und Sicherheitstechnik, Bergische Universität Wuppertal, Wuppertal, Germany}
\affil[5]{Institut für Theoretische Physik, Universität zu Köln, Köln, Germany}

\affil[*]{Corresponding authors: 
Oscar Dufour (oscar.dufour@univ-lyon1.fr),
Mohcine Chraibi (m.chraibi@fz-juelich.de),
Benoit Gaudou (benoit.gaudou@ut-capitole.fr),
Alexandre Nicolas (alexandre.nicolas@cnrs.fr),
Antoine Tordeux (tordeux@uni-wuppertal.de)}

\begin{abstract}

The dynamics of dense crowds have received considerable attention from researchers seeking
fundamental understanding or aiming to develop data-driven algorithms to predict pedestrian trajectories. However, current research mainly relies on data collected in controlled settings. We present one of the first comprehensive field datasets describing dense pedestrian dynamics at different scales, from contextualized macroscopic crowd flows over hundreds of meters to microscopic trajectories (around 7000 individual trajectories). In addition, a sample of GPS traces, some statistics of contacts and pushes, and a list of non-standard crowd phenomena observed in the video recordings are provided. Data were collected during the 2022 Festival of Lights in Lyon, France, as part of the French-German MADRAS project, and cover densities up to 4 pedestrians per square meter.  We suggest using this extensive data set, acquired in complex real-world settings, to benchmark models of pedestrian dynamics.

\end{abstract}

\begin{document}

\flushbottom
\maketitle

\thispagestyle{empty}


\tableofcontents

\newpage
\section{Background \& Summary}


\blue{
}

Large gatherings raise challenges regarding public safety and flow management; these challenges tend to be all the more acute as the crowd is large and dense. Religious festivals, music concerts, and significant outdoor events are therefore of particular concern, with a record of poor crowd management and, in the worst cases, deadly crowd crushes \cite{helbing2007dynamics,sieben2023inside,feliciani2023trends,sharma2023global}. Beyond the rules of thumb that have been refined over the years, a deeper fundamental understanding of the dynamics of dense crowds will be instrumental for more efficient event planning and crowd management. Here, a crowd will be described as dense if its density exceeds the arbitrary threshold of $1.5$ or $2\,\mathrm{ped/m}^2$, thus falling in Fruin's Level of Service F \cite{fruin1970designing}; critical conditions with extreme densities (above $8 \mathrm{ped/m}^2$), which must be avoided in practice, are left out of our scope.

The present theoretical grasp of dense crowd dynamics mainly stems from controlled experiments \cite{haghani2020empiricalI}, conducted in idealized settings. This protocol enables researchers to finely control variables and settings to observe a whole gamut of none-too-common scenarios, such as emergency evacuations \cite{garcimartin2018redefining} or high-density flows and crossings. The Research Center of J{\"u}lich, in particular, has collected numerous pedestrian trajectories in a wide array of such experiments \cite{zhang2014comparison,cao2017fundamental}. Let us also mention, among others, the datasets collected by the groups of Haghani \cite{haghani2020evacuation} and Zuriguel \cite{pastor2015experimental} to explore crowd dynamics during emergency evacuation drills, Murakami et al. \cite{murakami2021mutual} to probe the emergence of unidirectional and bidirectional pedestrian flows. An open-access data archive can be found here \cite{ExpJuelich}. However, the controlled conditions may substantially differ from reality, and the participants are aware of their involvement in the research. Another drawback of controlled experiments is that they do not afford a comprehensive picture of the crowd; the broader context in which the dynamics occur is missing.

The thirst for \emph{field data} to train data-based methods, such as machine learning algorithms, remains unquenched for dense crowds: field studies typically involve situations of low density. For instance, the widely used ETH \cite{pellegrini2009you} and UCY \cite{lerner2007crowds} datasets, which originate from surveillance videos, capture pedestrian scenes at density $0.1-0.5 \mathrm{ped/m}^2$, with many avoidance situations. In this regime, the pedestrian dynamics are believed to be governed by different mechanics than at higher density \cite{fruin1970designing,best2014densesense,cordes2023dimensionless}. Empirical datasets often encompass a heterogeneous mix of road users, as in the Stanford Drone Dataset (SDD) \cite{robicquet2016learning}, including pedestrians, cyclists, skateboarders, cars, and buses. While these datasets capture small scenes, the Grand Central Station Dataset (GS) \cite{zhou2012understanding}, collected in New York, covers a vast area. Although one scene can contain hundreds of pedestrians, the average density is below $0.2 \mathrm{ped/m}^2$ due to the region's size.
The Edinburgh Informatics Forum Dataset (EIF) \cite{majecka2009statistical} provides trajectories of $92\,000$ people in a university playground from an overhead camera, with minute average densities (most of the time just a few pedestrians at a time). Also worth mentioning are mass-gathering studies relying on a sparse sample of smartphone signals \cite{wirz2013probing}. 
Also worth mentioning is the Technical University of Eindhoven and the group of Toschi and Corbetta, which has been collecting extensive datasets of pedestrian trajectories in the Eindhoven train station and on the university campus in the last decade \cite{corbetta2014high,corbetta2020high}. Several reviews on existing field studies of pedestrian trajectories have recently appeared\cite{amirian2020opentraj,haghani2020empiricalII,korbmacher2022}. 

The present contribution aims to make up for the lack of field data on dense crowds by providing a comprehensive picture of pedestrian flows at a large gathering around a hot spot of a major cultural and entertainment event, the 2022 edition of the Festival of Lights, which took place in Lyon, France. For this purpose, we collected various data relevant to pedestrian dynamics and crowd management. These data cover an extensive range of length scales, by all standards in the field, from the global flow picture and contextual elements down to individual pedestrian trajectories and some statistics on physical contacts. In the following, an emphasis shall be placed on all observations that depart from what is typically prescribed or observed in controlled experiments, thus further highlighting the added value of actual field data.

Lyon's yearly Festival of Lights is a four-evening event (from December 7 to 11 in 2022, mostly from 7 pm to 11 pm) wherein the city is lit up remarkably. Originally a religious tribute to the Virgin Mary, it has become a massive international festival renowned for its innovative light shows and artistic projections on historic buildings. The event attracts millions of local and international visitors (more than 2 million officially in 2022\cite{LeProgres2022}). Key attractions include \emph{Place des Terreaux} and \emph{Place Saint-Jean}, reportedly attended by $150\,000$ and $80\,000$ spectators every night, respectively, in 2022\cite{LeProgres2022}. Quite interestingly, for our purposes, managing the associated crowd flows is one of the most prickly issues for the event organizers\cite{ActuLyon2023}. They aim to ensure smooth flows and reasonable delays for a pleasant experience. Still, above all, to ward off crowd accidents, after a difficult situation witnessed in the 2000s\footnote{Private communication with the organizers of the event.} and the tragedies that have occurred in massive entertainment events around the world, e.g. at the Love Parade in Duisburg, Germany, in 2010\cite{sieben2023inside} or during Halloween on the streets Seoul, Korea, in 2022\cite{liang2024unraveling}. This is achieved by regulating flows at different scales: macroscopically, by suggesting routes through the city to visit the multiple light installations, for instance, starting at \emph{Place Bellecour}, moving to \emph{Place des Terreaux}, and then exploring \emph{Vieux Lyon}, especially near \emph{Saint-Jean's Cathedral}; mesoscopically, by installing barriers and safety agents, particularly near \emph{Place des Terreaux} and imposing unidirectional flows in many streets to ease congestion microscopically by continuously monitoring the event with CCTV. Our methodology explores these three scales for crowd flows, focusing on the microscopic one at the central location of \emph{Place des Terreaux}. The crowd's movement was notably monitored with strategically placed cameras (Fig. \ref{fig:outflow_extraction}) installed by us or by the city of Lyon.


\section{Methods}
\label{sec:methods}

The datasets \cite{data_madras_Geometry,data_madras_GPS,data_madras_LargeViewTrajectories,data_madras_surveys,data_madras_TopViewTrajectories} referenced in this study in relation with dense pedestrian dynamics during the 2022 Festival of Lights in Lyon, France, are publicly available and can be accessed via Zenodo\cite{data_madras_Geometry,data_madras_GPS,data_madras_LargeViewTrajectories,data_madras_surveys,data_madras_TopViewTrajectories}. The datasets include macroscopic crowd flows, microscopic individual trajectories, GPS traces, and statistics on crowd interactions, providing a valuable resource for further research in pedestrian dynamics.
\subsection*{General organization of the data collection campaign}

Approximately ten staff members planned and carried out the data collection campaign. The aim was to capture a complete picture of pedestrian motion during the Festival of Lights. For that purpose, we combined various methods targeting different scales. 
Macroscopically, we inspected the broad patterns of macroscopic crowd flows on the ground around \emph{Place des Terreaux} and recorded the scene with a broad overview. We surveyed passers-by close to the entrance of the square, asking them the following questions:

\begin{itemize}[nosep]
    \item `How many people were with you?' 
    \item `How many children were with you?'
    \item `What was the last screening you attended before the one at Place des Terreaux?'
    \item `What screening do you plan to attend next? (You can answer that you don't know)'
\end{itemize}  
We strove to limit the selection bias on the the $79$ respondents by making no distinction about general or physical appearance and by ensuring that we interviewed a sample from around the square and not limited to a single street. 

Mesoscopically, we recruited participants willing to share their GPS data and record how often they bumped into other people.

\begin{figure}[htbp]
    \centering
    \includegraphics[width=0.7\textwidth]{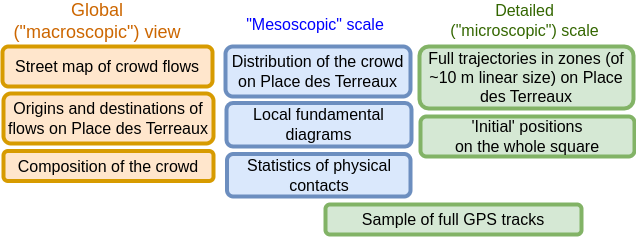} 
    \caption{Scales probed in this study and synthesis of the data collected at each scale.}
    \label{fig:flow_chart}
\end{figure}

Microscopically, we installed several cameras filming specific zones from the top (with all required authorizations to ensure privacy preservation). Below, we make a distinction between \emph{TopView} cameras (labelled 1, 2, 3, 5, 6, 7, 8) and \emph{LargeView} cameras (labelled 4).
An overview of the data collected in this study is presented in Fig.~\ref{fig:flow_chart}.

\subsection*{GPS data of recruited participants and collision counts}

A group of 24 undergraduate and doctoral students from \emph{Institut Lumière Matière} in Lyon were invited to participate in the field study. More precisely, they were asked to follow a general route (from the queue at the entrance to the square, on \emph{Rue du Président Edouard Herriot} to the exit on \emph{Rue Lanterne}, following the flow of the crowd), behave as a standard spectator. Meanwhile, their GPS positions were recorded on their smartphones using the \emph{GeoTracker} application. Although the accuracy of the measurements varied from phone to phone, the trajectories obtained were generally reliable, with a margin of error of around 10 meters on absolute positions. Besides, each participant used a stopwatch on their smartphone to record each time they collided with another pedestrian. Upon synchronization with the GPS data, this method allowed us to precisely determine the time and location of the collisions. Each participant was taught what to consider a collision, which was demonstrated by jostling them hard. Therefore, minor rubbing of the clothing was not recorded. Of the 24 students involved in that experiment, 16 were able to provide us with a detailed list of physical contacts, among which eight could be coupled to GPS data (Dataset~\cite{data_madras_GPS}).
Our efforts to involve on-site spectators in the campaign proved fruitless; none managed to send us GPS data.

\begin{figure}[htbp]
    \centering
    \includegraphics[width=\linewidth]{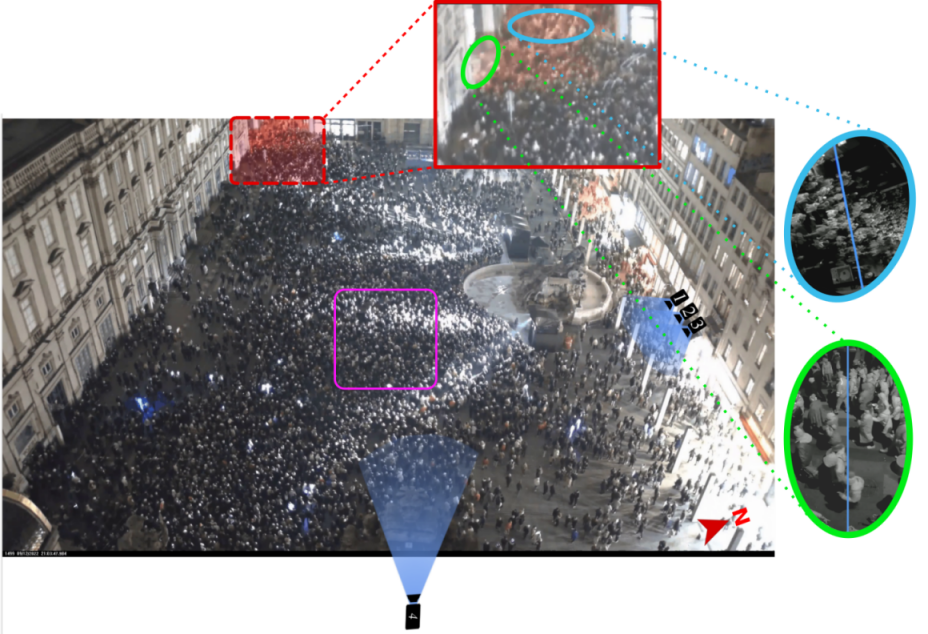}\,
    \caption{The thick crowd gathered in and around \emph{Place des Terreaux} at the end of a light show on December 12, $2022$, at $9$:$41$ PM. The fields of view of the cameras are highlighted in blue; the pink rounded square delimits the area where pedestrians were exhaustively tracked in the \emph{large-view} datasets\cite{data_madras_LargeViewTrajectories} (\url{https://www.youtube.com/watch?v=jIC3dNOZsk0}). Inset: Zoom on two exit streets where we monitored the outflow.}
    \label{fig:outflow_extraction}
\end{figure}

\subsection*{Video processing and pedestrian tracking for the \emph{TopView} cameras}
\label{sub:methods_TopView}

To obtain a finer view of pedestrian trajectories, we deployed lightweight cameras \emph{SJCAM A10} in strategic locations around \emph{Place des Terreaux}. These cameras filmed the scene from a zenith perspective. These cameras were selected for their night vision capabilities and long battery life; three of them, labeled $1$, $2$, and $3$ in Fig.~\ref{fig:maps}, were placed on the north side of \emph{Place des Terreaux}, protruding from the windows of an apartment \emph{Airbnb} and a restaurant on the second floor. They captured the bidirectional movement of pedestrians below. Another camera, numbered $8$, was mounted atop an existing post in the square's South-East corner to monitor incoming flows; however, its nighttime footage was unusable due to lighting problems. Two additional cameras, numbered $5$ and $6$, were placed on \emph{Rue Constantine}, one of the main exits after the light show, hanging from the balconies of an \emph{Airbnb} apartment to film the pedestrian egress from above. A final camera numbered $7$ was temporarily held at the end of a stick in the South-West corner to provide a closer view of the pedestrian outflow. Overall, the portable \emph{SJCAM A10} cameras recorded nearly $200$ GB of video at $30$ frames per second.

From the collected footage, pedestrian trajectories were extracted from 10 excerpts (see Table~\ref{tab:StatAirBB} and Dataset~\cite{data_madras_TopViewTrajectories}), using the \emph{PeTrack} software \cite{boltesCollectingPedestrianTrajectories2013, boltes_PeTrack_2022}. This software is commonly used to automatically detect and track pedestrian heads in controlled experimental settings. 
The process notably involves calibrating the cameras to match pixel coordinates with real-world coordinates, which is split into two distinct phases: internal and external calibration. Internal calibration corrects the specific optical distortions of each camera lens by determining the optimum parameters to transform a reference pattern (such as a checkerboard) into its recorded image. 
Then, external calibration yields the conversion between real-world coordinates and pixel coordinates using three successive operations: a rotation and a translation of the reference frame, followed by a projection to move from the camera frame to the screen frame. The parameters of these operations minimize the differences between the known real-world coordinates of objects (here, staff members) positioned at predefined positions (here, at regularly spaced positions on a virtual `grid') and the associated pixel coordinates. 
Such calibration is only an approximation if the recorded pedestrians are of unequal heights, especially without a stereoscopic camera to reconstruct the scene in three dimensions. 
Still, the inaccuracy is even lower as the camera is positioned higher and films from a zenithal viewpoint. 
Here, we expect the maximum uncertainty due to unequal heights to be of order\cite{boltes2010automatic} $\delta h \cdot \frac{D}{H}$, where $H\approx 12\,\mathrm{m}$ is the altitude of the camera concerning the heads, $D \approx 5-10\,\mathrm{m}$ is the maximal horizontal stretch between the camera axis and people on the periphery of the scene, and $\delta h \lesssim 20\,\mathrm{cm}$ is the height difference between filmed people and the staff member who served as reference for the calibration; hence below $15\,\mathrm{cm}$ in most cases for deployed \emph{SJCAM A10} cameras. The uncertainty in detecting the central point on the head ($\lesssim 5 \,\mathrm{cm}$) and the error should be added to this because not all people stand perfectly upright. 
Furthermore, occlusion phenomena due to the perspective can increase the measurement uncertainty for small pedestrians.

After the calibration step, the \emph{PeTrack} software semi-automatically detects and tracks pedestrians on the videos. Compared to controlled experimental conditions, some difficulties arose. 
In our case, this task was mainly complicated by the lighting conditions. 
Thus, pedestrian heads were first detected manually and then tracked from frame to frame using the extended pyramidal iterative Lucas Kanade feature tracker integrated into \emph{PeTrack}. Although this method is surprisingly robust, manual corrections were often required, mainly when the illumination suddenly changed from dark to bright areas.


\subsection*{Video processing and pedestrian tracking for the \emph{LargeView} camera}
 
In addition to the previously mentioned \emph{TopView} cameras, we have gathered extensive \emph{large-view} video footage from cameras that offer a bird's-eye view of the entire square. Two of these cameras are located at \emph{Place Saint-Jean}. Another camera (numbered $4$ in Fig.~\ref{fig:maps}) captures \emph{Place des Terreaux} from the City Hall tower, approximately $70$ meters above ground level. Although all recordings are accessible, only the videos from \emph{Place des Terreaux} have been thoroughly analyzed and discussed here.
 

The expansive view provided by the camera, numbered $4$, along with variations in lighting, precludes automated tracking using \emph{PeTrack}, in favor of a largely manual approach. Internal calibration was deemed unnecessary because the camera lens exhibited minimal optical distortion: straight lines in reality appear as straight lines on the video. Complete external calibration was performed by positioning a staff member at predefined evenly spaced points in the square $32$ and applying the geometric transformations outlined in Sect.~\ref{sub:methods_TopView}. The resulting precision for \emph{absolute} positions, assessed using independently collected positions (either of the same staff member or others), ranged from $10$ cm to approximately $2$ meters at the farthest end of the square and over $80$ meters from the camera horizontally. Given that this inaccuracy is primarily geometric, it is expected that the positions \emph{relative} between pedestrians and their neighbours are significantly more precise.

Initially, we manually identified the positions of all individuals throughout the square at a specific time point, occasionally reviewing video frames to locate pedestrians temporarily obscured from view. This was done after a show cycle, as people began to exit, on Thursday 8 December 2022. Subsequently, we tracked a random sample of approximately $270$ individuals over several seconds, with around $100$ of them being tracked for a duration of $20$ seconds (Dataset~\cite{data_madras_LargeViewTrajectories}, file "LargeView\_tracers.txt").
Finally, we focus on a particular area of interest, where opposing flows of people who do not necessarily follow traffic directions meet. The area has a square shape (before the correction of the geometric distortion) and is located near the fountain, a crucial convergence point of high density. Using homemade Python-based software, we manually tracked the trajectories of all observable people in this area over around 30 seconds, at a rate of typically two frames per second. Trajectories were upsampled to 10 Hz by linear interpolation and exported into CSV files (dataset~\cite{data_madras_LargeViewTrajectories}, files "LargeView\_zoom\_A.txt" and "LargeView\_zoom\_O.txt").

It should be noted that, compared to the \emph{TopView} videos, these ones (addressing a more complex flow scenario) are of lower resolution and quality. Combined with the varying illumination, this hindered the detection of some individuals, especially children and shorter people, in specific frames. The datasets may thus exhibit imperfections, such as some individuals escaping detection and occasional swaps between trajectories. However, our subsequent tests have shown that they are accurate to a very large degree and nearly comprehensive (see \emph{Technical validation}).

\subsection*{Conversion into global coordinates}

All trajectories were mapped to global coordinates using the RGF-93 Lambert-93 coordinate reference system (EPSG:2154) for precise global positioning. This transformation involved adjusting the positions of several landmarks visible in the videos and satellite imagery. The geometric shapes and locations of all obstacles in the square, including the fountain, temporary crowd barriers surrounding it, bollards, and Buren columns, were determined via direct measurements, together with satellite imagery and photographs taken on-site; they are provided as an external dataset\cite{data_madras_Geometry}.

\subsection*{Quantitative indicators}

Various relevant static and dynamic indicators can be computed using the detailed pedestrian trajectories. 

\textbf{\textit{Flow rate.}}  The outflow rate during an egress from \emph{Place des Terreaux} was measured by drawing virtual cross-section lines at the two main exits, as shown in Fig.~\ref{fig:outflow_extraction}. People crossing these lines were counted manually  by watching the recorded videos.

\textbf{\textit{Density field.}}  To be useful, the extracted microscopic trajectories often need to be smoothed into continuous fields. Specifically, the local density field\cite{johansson2008crowd}, denoted as $\rho(\mathbf{r}, t)$, is derived by computing the convolution of the microscopic particle density $\rho_{\mu}(\mathbf{r}, t) = \frac{1}{A} \sum_j \delta \left(\mathbf{r} - \mathbf{r}_j(t)\right)$ — where $\delta(\cdot)$ is the Dirac delta function and $A$ is the surface area — with a Gaussian kernel $\phi_{\xi}(\mathbf{r}) \propto \exp\left(-\frac{\mathbf{r}^2}{2\xi^2}\right)$ whose integral is normalized to 1, for a chosen half-width $\xi$, viz.:
\begin{equation}
    \rho(\mathbf{r}, t) = \int_{\mathcal{A}} \rho_{\mu}(\mathbf{r} - \mathbf{r'}) ~ \phi_{\xi}(\mathbf{r'}) ~ \mathrm{d}^2\mathbf{r'}
\end{equation}
For further smoothing, the time dependence can also be coarse-grained by averaging the field over a short time window $\Delta t$: 
\begin{equation}
    \bar{\rho}(\mathbf{r}, t) = \frac{1}{\Delta t} \int_{t - \frac{\Delta t}{2}}^{t + \frac{\Delta t}{2}} \rho(\mathbf{r}, t') \, \mathrm{d}t'
\end{equation}


\textbf{\textit{Velocity field.}} Similarly, the microscopic velocities $\mathbf{v}_j(t)=\frac{1}{\delta t} \cdot \Big[ \mathbf{\tilde{r}}_j(t+\delta t) - \mathbf{\tilde{r}}_j(t)\Big]$, estimated from the vector difference
between two positions of pedestrian $j$ over a small time interval $\delta t$ (with $\delta t = 0.5$ or $1$ second in this context), can be coarse-grained. The trajectories are initially smoothed using a second-order Butterworth low-pass filter. Trajectories that are too short for effective filtering remain unfiltered. Near the start ($t_s$) and end ($t_e$) of each trajectory, linear interpolation is applied between the raw trajectory $\mathbf{r}$ and the filtered trajectory $\mathbf{\tilde{r}}$ to address the Butterworth filter's limitations when past or future data points are absent. This interpolation is expressed as\footnote{At the start ($t = t_s$) and end ($t = t_e$) of the trajectory, $\alpha$ is close to $1$, giving more weight to the original data. This helps preserve the trajectory's endpoints. In the middle, $\alpha$ decreases, giving more weight to the smoothed data. The beginning and end of the raw trajectories are thus preserved, reducing artifacts from filtering.}
 $\mathbf{\tilde{r}}_j(t) \leftarrow
 \alpha(t)\mathbf{r}_j(t)  + [1-\alpha(t)]\mathbf{\tilde{r}}_j(t)$, where $\alpha(t)=\mathrm{max}\left\{ e^{-(t-t_s)},\ e^{-(t_e-t)} \right\}$.
Subsequently, the trajectories are transformed into a velocity field through Gaussian convolution:
\begin{equation}
\label{eq:GaussianKernel}
    \mathbf{v}(\mathbf{r}, t) = \frac{\sum_j \mathbf{v}_j(t) \, \phi_{\xi}\left(\mathbf{r} - \mathbf{\tilde r}_j(t)\right)}{\sum_j \phi_{\xi}\left(\mathbf{r} - \mathbf{\tilde r}_j(t)\right)}
\end{equation}
The resulting field exhibits abrupt variations due to many individuals deviating from, or walking counter to, the primary local flow direction. Like the density field, the velocity field can be averaged over a finite time interval $\Delta t$ to smooth out these variations:
\begin{equation}
\bar{\mathbf{v}}(\mathbf{r}, t) = \frac{1}{\Delta t} \int_{t - \frac{\Delta t}{2}}^{t + \frac{\Delta t}{2}} \mathbf{v}(\mathbf{r}, t') \, \mathrm{d}t'
\end{equation}
The coarse-grained picture given by the smooth velocity field masks possible counterflows and fluctuations, whose presence can be ascertained by computing a \emph{variance} field:
\begin{equation}
\mathrm{Var}_{\mathbf{v}}(\mathbf{r}, t) =  
\frac{ 
    \int_{t - \frac{\Delta t}{2}}^{t + \frac{\Delta t}{2}} \sum_j \phi_{\xi}\left(\mathbf{r} - \mathbf{r}_j(t')\right)  
    \left\| \mathbf{v}_j(t') - \bar{\mathbf{v}}(\mathbf{r}, t') \right\|^2 \, \mathrm{d}t' 
}
{ 
    \int_{t - \frac{\Delta t}{2}}^{t + \frac{\Delta t}{2}} \sum_j \phi_{\xi}\left(\mathbf{r} - \mathbf{r}_j(t')\right) \, \mathrm{d}t'
}
\end{equation}
\noindent Trajectories of counter-walking pedestrians $j$ significantly deviate from the continuous flow, thus exhibiting a large variance relative to the velocity field, denoted as $\mathrm{Var}_{\mathbf{v}}^{j}$. More precisely, the variance is computed by averaging the squared difference between pedestrian $j$'s velocity 
$\bs{v}_j(t)$ and the coarse-grained velocity $\bs{\bar{v}}(\bs{r}_j(t),t)$ at that position, over the entire duration of pedestrian $j$'s trajectory, i.e.
$\mathrm{Var}_{\bs{v}}^{j}= \langle \left\| \bs{v}_j(t) - \bs{\bar{v}}(\bs{r}_j(t),t) \right\|^2  \rangle$.

\textbf{\textit{Fundamental diagram.}} The fundamental diagram is obtained by relating the instantaneous pedestrian speeds 
$\left\| \bs{v}_{j}(t) \right\|$
to the local density $\rho$ (binned in cells of linear size $0.25\,\mathrm{m}$ and duration $0.5\,\mathrm{s}$).

 For more information on defining quantitative indicators, refer to the documentation of the MADRAS-Streamlit\cite{streamlit} and PedPy\cite{pedpy} libraries.



\section{Data Records}

\subsection{Composition and size of the crowd; macroscopic flows directions}

\begin{figure}[!ht]
    \centering
    \includegraphics[width=\textwidth]{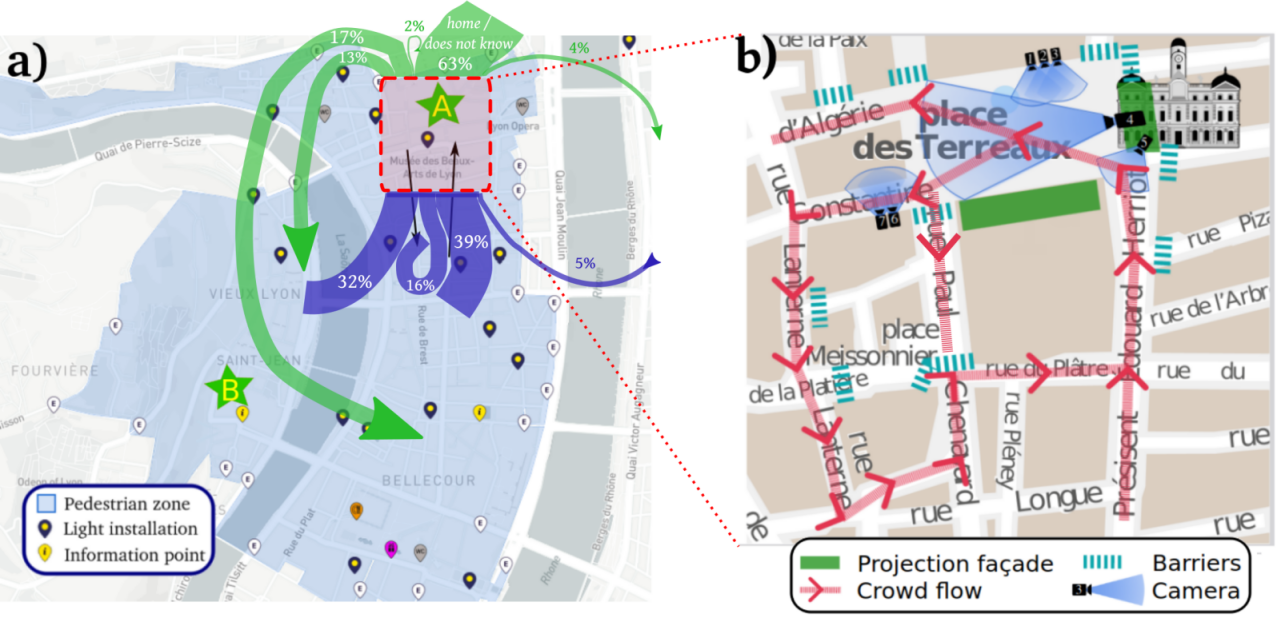} 
    \caption{Maps of the macroscopic crowd flows during the Festival of Lights. \textbf{(a)} Pedestrian zone 
    adapted from the official map \cite{OfficialMap} 
    indicating the distribution of origins (in blue) and destinations (in green) of around 300 people, just before and just after the show on \emph{Place des Terreaux}, obtained by surveying 79 passers-by around 11 pm on Friday 9 December 2022. \textbf{(b)} Local map showing the imposed flow directions around \emph{Place des Terreaux}. All figures are oriented to the North unless otherwise shown in the figure.}
    \label{fig:maps}
\end{figure}

\emph{Place des Terreaux} is centrally located in Lyon and is a key attraction during the Festival of Lights. Pedestrian traffic around the square is regulated: spectators enter from the South-East via \emph{Rue du Président Edouard Herriot}, typically remain in the square for the duration of one show (6 minutes and 30 seconds), sometimes two, and then exit either to the South-West via \emph{Rue Constantine} or to the North-West via \emph{Rue d'Algérie}, as illustrated in Fig.~\ref{fig:maps}b.

\begin{figure}[!ht]
    \centering
    \includegraphics[width=0.5\linewidth]{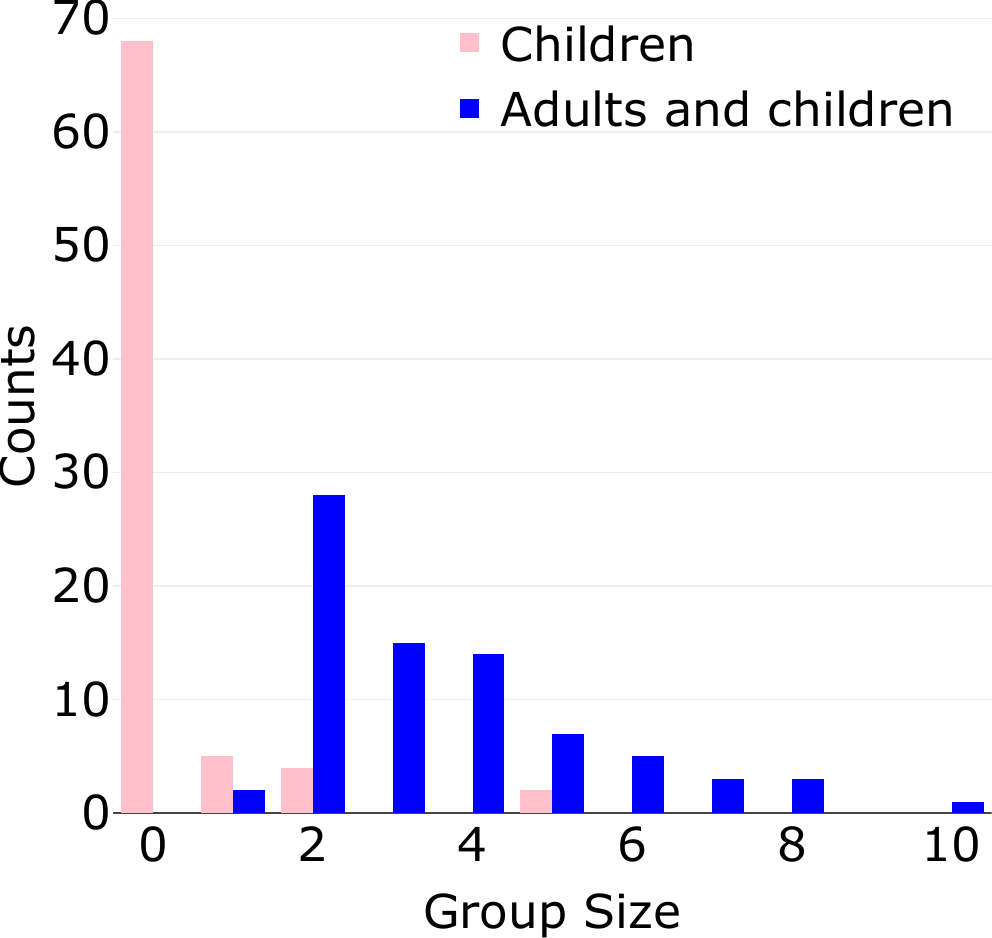}\vspace{-2mm}
    \caption{Histogram of social group sizes, counting both adults and children (in blue) or only the children (in light red), as reported by survey respondents around 10:30 pm to 11 pm on Friday December 9, 2022. These groups were part of the crowd preparing to enter \emph{Place des Terreaux}. The counts refer to the number of groups. It is important to note that these stated values (survey) contain a significantly higher number of groups with at least four members compared to estimates made by the authors by means of direct street observations (Dataset~ \cite{data_madras_surveys}).
    }
    \label{fig:group_sizes}
\end{figure}

Field surveys conducted around 11 pm on Friday revealed that most spectators had previously seen light animations just south of the square. Upon entering \emph{Place des Terreaux}, many were uncertain about their next destination or planned to head home. These large-scale origin-destination flows are depicted in Fig.~\ref{fig:maps}a. Spectators usually belong to social groups of two to four people; larger groups, up to ten people (and above), also exist, but become less frequent as the group size increases. Most groups do not include children, although groups with one or two children were observed several times, as shown in Fig.~\ref{fig:group_sizes}.

The entrance to \emph{Place des Terreaux} is managed by gatekeepers who ensure that the square does not exceed approximately two-third of its maximum capacity. They restrict access by closing a barrier before the start of each show, leading to a queue of people standing on \emph{Rue du Président Edouard Herriot} that can stretch over several blocks. During a light show, the number of people in \emph{Place des Terreaux} can significantly exceed $4\,000$, according to our 
manual detections and counts on one snapshot extracted from the videos (refer to Table~\ref{tab:CCTVstats}). 

\begin{figure}[!ht]
    \centering
    \begin{subfigure}[t]{0.45\textwidth}
        \centering
        \includegraphics[width=0.9\textwidth]{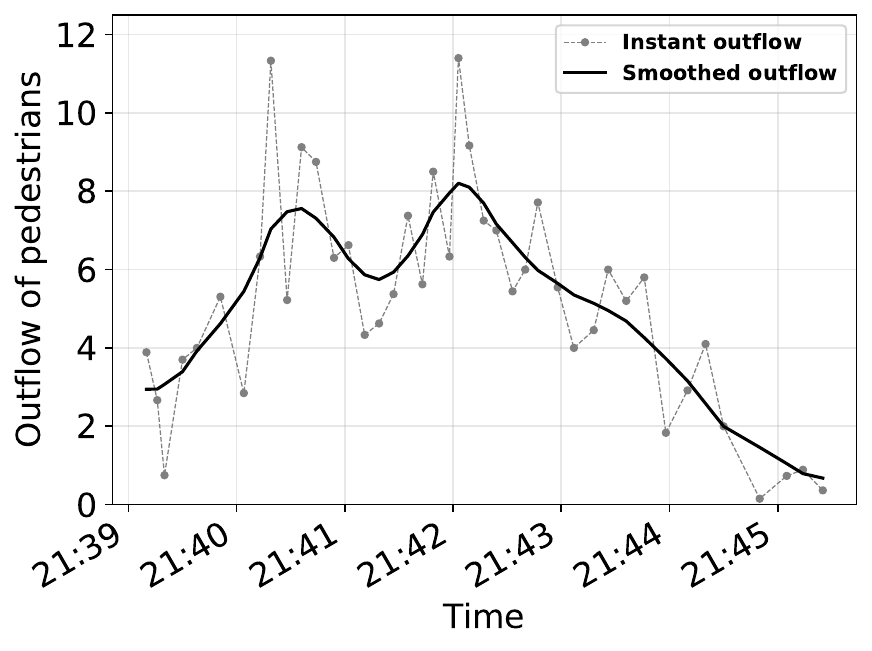}
        \caption[1]{\emph{Rue Constantine}}
    \end{subfigure}%
    ~ 
    \begin{subfigure}[t]{0.45\textwidth}
        \centering
        \includegraphics[width=0.9\textwidth]{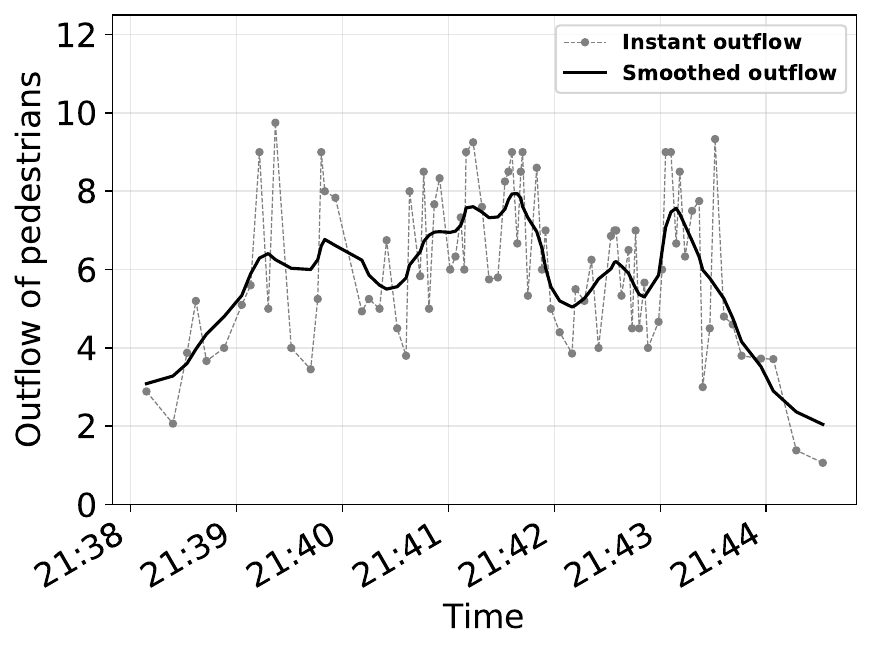}
        \caption{\emph{Rue Paul Chenavard}}
    \end{subfigure}
    \caption{Pedestrian outflows measured during one of the periodic egresses from \emph{Place des Terreaux} on December 10, 2022.}
    \label{fig:pedestrian_outflow_results}
\end{figure}

This attendance number is corroborated by the cumulative pedestrian outflow measured at the two exits, \emph{Rue Constantine} and \emph{Rue Paul Chenavard}, at the end of a show, from 9:38 pm to 9:45 pm on December 9, 2022 (see Sec.~\ref{sec:methods}); the total evacuation time was approximately 6.5 minutes.
The evolution of pedestrian outflows over time is shown in Fig.~\ref{fig:pedestrian_outflow_results}. It displays the raw, instantaneous values (depicted by dashed black lines) and the smoother curves (solid black line) obtained by applying a Gaussian filter with a kernel standard deviation of 2.0. The maximum outflow exceeded 11~ped/s on \emph{Rue Constantine} and nearly reached 10~ped/s on \emph{Rue Paul Chenavard}. In total, $3833$ pedestrians were counted, with $1803$ on \emph{Rue Constantine} and $2030$ on \emph{Rue Paul Chenavard}. 

Around $4\,000$ people on the whole square correspond to a global density below one pedestrian per square meter. However, as we will observe, the global average is not particularly insightful due to significant spatial heterogeneity. It is important to note that both counting methods tend to underestimate the actual numbers: on a snapshot, not all individuals are clearly visible, while the estimate based on the outflows discards the spectators who did not leave the square after the show. Besides, despite the regulated inflow, variations in the number of attendees occur during each cycle. 
However, quite remarkably, visual inspection of the entrance and egress flows points to a high degree of regularity of the flows across cycles. In particular, obstacles and prohibited areas (marked in blue in Fig.~\ref{fig:PlTerreaux_full}) visibly alter the flow pattern, creating a confluence zone between the fountain and the northern building, along with congestion points on the opposite side of the fountain. In contrast, the crowd moves almost freely along the southern side, where projections are displayed on a building during the show, thus making this viewpoint less attractive to spectators (see Fig.~\ref{fig:outflow_extraction}).

\begin{figure}[!ht]
\centering
\includegraphics[width=\linewidth]{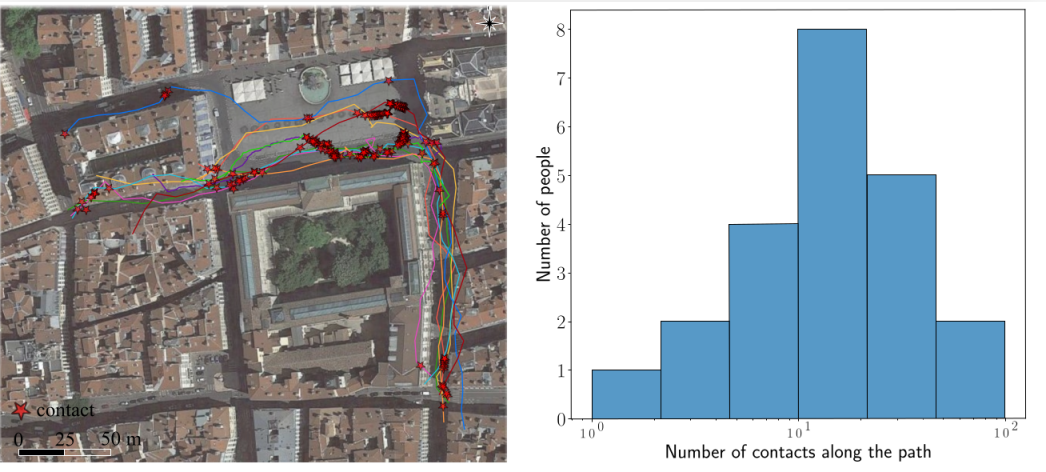}
\caption{\textbf{Left panel:} Ten GPS tracks are paired with contact data, with each contact marked by a red star. Clusters of red stars appear at the entrance and exit of \emph{Place des Terreaux}. Note that the satellite view timing does not match the data, and some GPS tracks overlap buildings due to varying GPS precision across devices, providing a general idea of contact locations. \textbf{Right panel:} A histogram showing the distribution of contact counts along the path on a logarithmic scale, incorporating data from the left panel and additional entries without GPS tracks.}
\label{fig:GPS_contacts}
\end{figure}

The GPS trajectories collected from informed participants (Dataset~\cite{data_madras_GPS}), as depicted in Fig.~\ref{fig:GPS_contacts}, illustrate potential routes from the square's entrance to its exit and beyond. By synchronizing these trajectories with the reported times of pushes and strong contacts, we  identified where these interactions occurred and marked their locations as stars on Fig.~\ref{fig:GPS_contacts}. Conspicuously, the total numbers of reported pushes vary significantly among participants, ranging from nearly zero to approximately 100 throughout the trajectory, as noticeable in Fig.~\ref{fig:GPS_contacts}. These variations highlight the heterogeneity of the crowd packing, the diversity of individual behaviors, and, plausibly, different appraisals of what should be counted as a push. 
Still, the order of magnitude of the frequency of strong contacts questions the collision-free
navigation hypothesis at the heart of some models based on velocity obstacles \cite{van2011reciprocal,karamouzas2017implicit}, but also, at the other pole, the strong role played by contact forces at densities below 4~ped/m$^2$ in other (typically force-based) force-based models.

\subsection{Global view of the flow patterns on \emph{Place des Terreaux}}

\begin{figure}[ht!]
    \centering
    \includegraphics[width=0.46\linewidth]{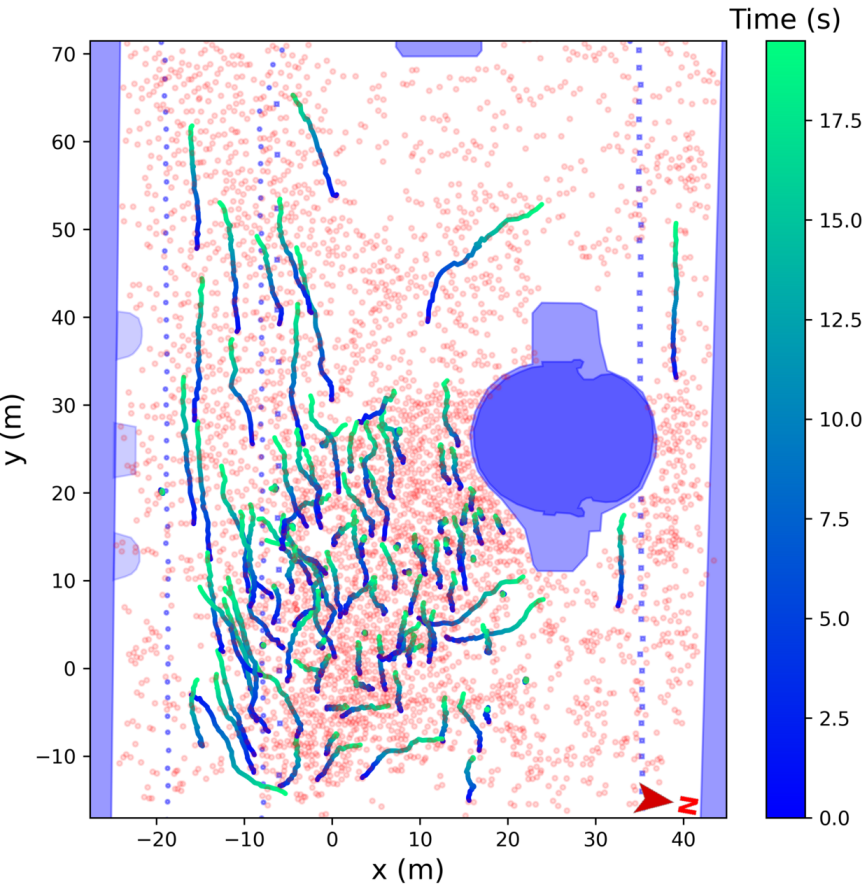}
    ~
    \includegraphics[width=0.48\linewidth]{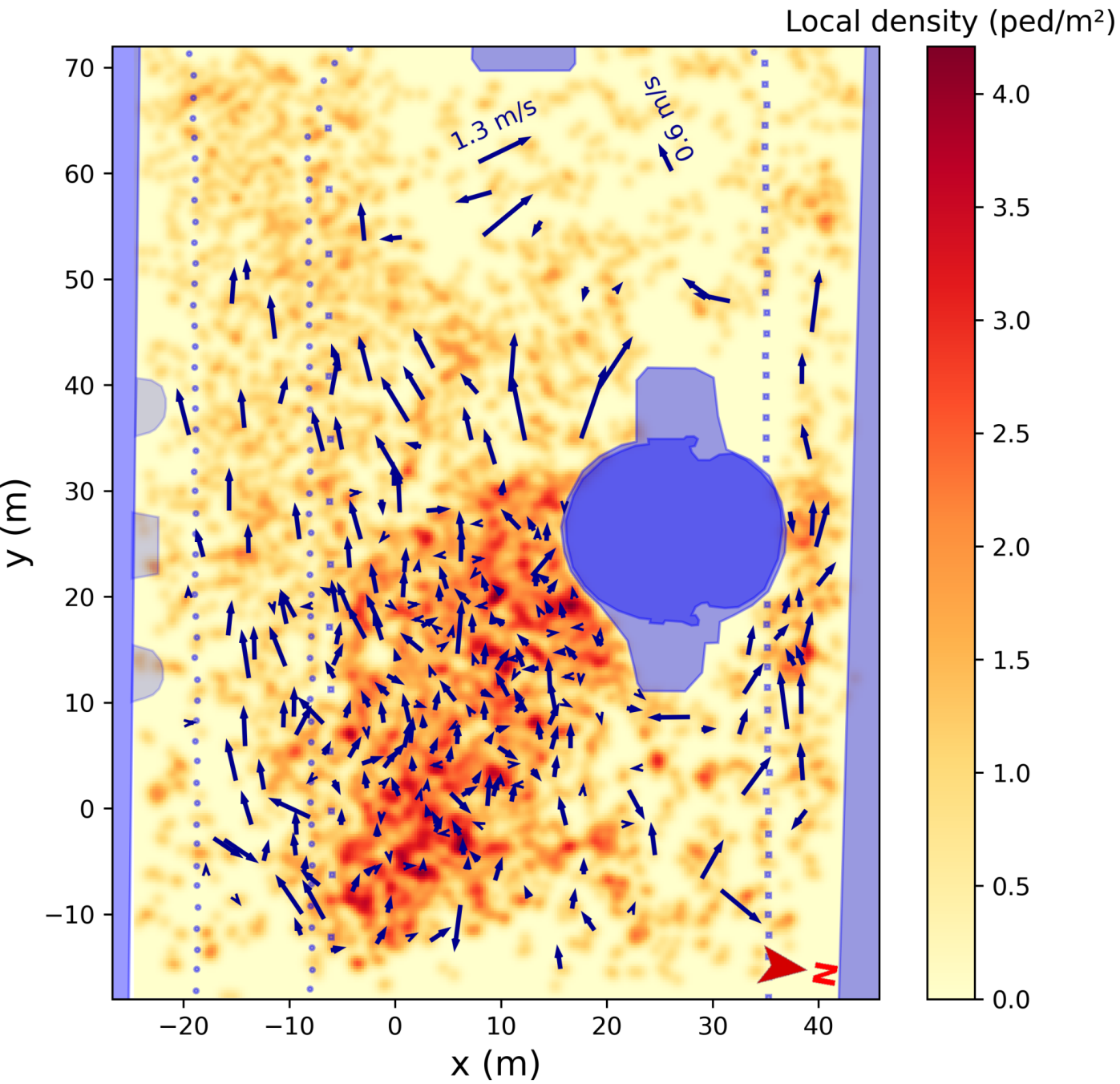}
    \caption{Comprehensive overview of the crowd's positions and dynamics on \emph{Place des Terreaux}. \textbf{Left Panel:} Trajectories of approximately $100$ pedestrians tracked over a span of $20$ seconds. Small red disks mark the initial positions of all individuals in the square. Various obstacles, such as fountains, buildings, bollards, and barriers, are shaded in blue (Dataset~\cite{data_madras_Geometry}). \textbf{Right Panel:} Initial velocities of around $270$ pedestrians, computed with a time step of $\Delta t=1\,\mathrm{s}$. The background features a heat map representing the initial density field, computed using a Gaussian kernel with a half-width of $\sigma=0.5\,\mathrm{m}$.}
    \label{fig:PlTerreaux_full}
\end{figure}

To better understand the distribution of the crowd and the flow patterns during the evacuation of the square after a show, we analyzed the \emph{LargeView} video recordings. From a snapshot captured at the start of the repeated evacuation process, we manually extracted the positions of all visible heads in the crowd. These positions are represented as small red disks in Fig.~\ref{fig:PlTerreaux_full}a, along with the accessible geometry of the square. The heterogeneous spatial distribution is manifest and becomes even more pronounced in the corresponding density field shown in Fig.~\ref{fig:PlTerreaux_full}b, where local densities range from nearly $0$ ped/m$^2$ to 4 ped/m$^2$. The trajectories of approximately $100$ randomly sampled pedestrians tracked over about $20$ seconds and shown in Fig.~\ref{fig:PlTerreaux_full}a (also see Supplementary Video \url{https://www.youtube.com/watch?v=jIC3dNOZsk0}), also exhibit marked heterogeneity, with a significant portion of pedestrians halted and some moving counter to the flow. To get a broader perspective on this heterogeneity, we measured the initial velocities of a larger sample of $270$ people over a few seconds, as shown in Fig.~\ref{fig:PlTerreaux_full}b (Dataset~\cite{data_madras_LargeViewTrajectories}). 

\begin{figure}[!ht]
    \centering
    \includegraphics[width=0.45\linewidth]{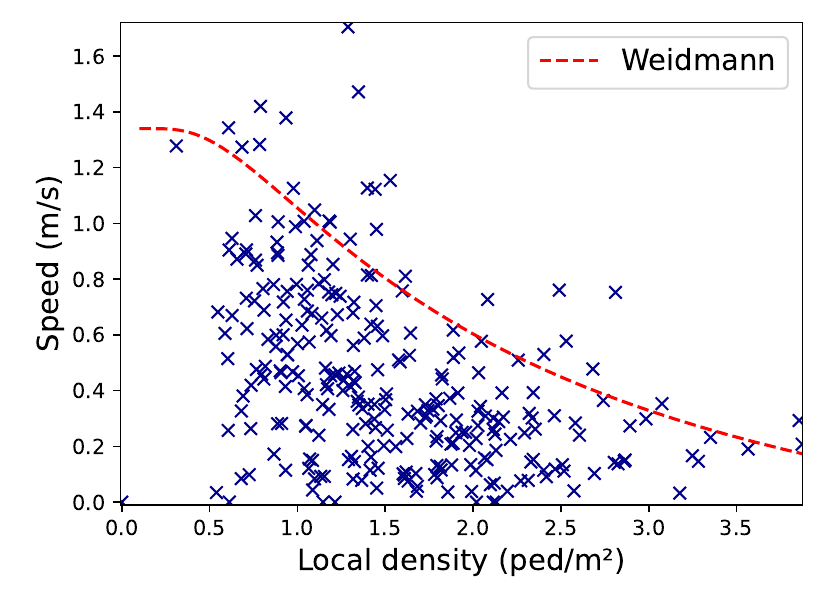}\vspace{5mm}
    \includegraphics[width=0.45\linewidth]{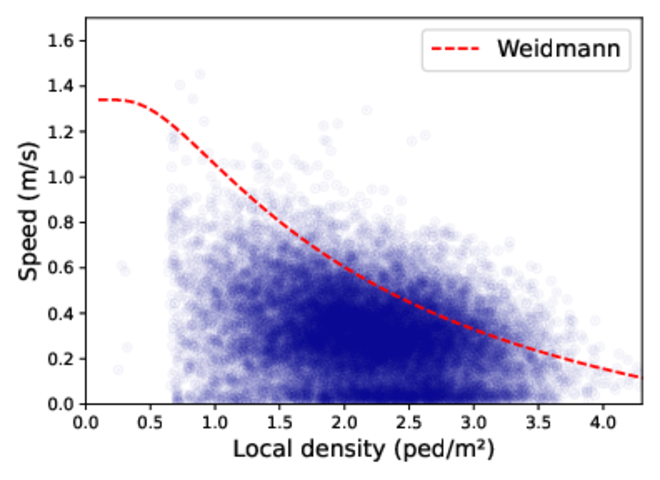}
    \caption{Fundamental diagrams relating pedestrian speed to the local density. The \textbf{left panel} depicts this relationship across the entire \emph{Place des Terreaux}, using a sample of initial velocities as shown in Fig.~\ref{fig:PlTerreaux_full} (right panel). The \textbf{right panel} focuses on a central region of the square, where pedestrians were exhaustively tracked for approximately $20$ seconds (see Fig.~\ref{fig:CCTV1}). Dashed red lines: Weidmann's empirical formula $v(\rho) = v_0 \cdot \left[ 1 - \exp\left(-\gamma \cdot \left(\frac{1}{\rho} - \frac{1}{\rho_{\mathrm{max}}}\right)\right) \right]$ with parameters $v_0 = 1.34\,\mathrm{m/s}$, $\gamma = 1.9\,\mathrm{m}^{-2}$, and $\rho_{\mathrm{max}} = 5.4\,\mathrm{ped/m}^2$.}
    \label{fig:PlTerreaux_FD}
\end{figure}

Plotting these results in terms of initial speed against local density provides the fundamental diagram presented in Fig.~\ref{fig:PlTerreaux_FD} (left).
Compared to conventional fundamental diagrams \cite{vanumu2017fundamental}, largely obtained in controlled settings, it exhibits much more scatter. In particular, while the largest speeds observed at a given density form an envelope curve broadly compatible with Weidmann's empirical formula \cite{wirz2013probing}
$v(\rho)= v_0 \cdot \Big[ 1 - \exp\Big(-\gamma\cdot(\frac{1}{\rho}-\frac{1}{\rho_{\mathrm{max}}}) \Big) \Big]$, many speed data points
fall between 0$\ms$ and this curve. They correspond to people strolling, sometimes because they are moving along (and possibly chatting) with a social group or who are halted. Interestingly, at a superficial level, this observation mirrors the doubts on the uniqueness (i.e., bijectivity) of the mapping between speed and density in vehicular traffic and the arguments in favor of a fundamental diagram spanning a two-dimensional region of the plane \cite{jiang2014traffic}.

That being said, general flow patterns are discernible during the evacuations of \emph{Place des Terreaux}. Most people head West, towards the two main exits (located at the top of the picture in Fig.~\ref{fig:PlTerreaux_full}). Between the fountain and the northern building (to the right of the picture), the flows (mainly, but not exclusively, directed to the West) go through a zone of convergence, which will be probed in greater detail below. On the road along the southern building, most trajectories stretch linearly, from East to West, in almost free-flowing conditions. In contrast, trajectories observed closer
to the fountain are significantly shorter over the same time interval, pointing to congestion and more diverse behaviors, with many people at a standstill. 

\begin{table}[ht]
    \centering
    \begin{tabular}{ p{2cm} p{2.2cm} p{1.5cm} p{2.7cm} p{3cm} p{2.5cm} }
        \hline
        \textbf{File} & \textbf{Date [UTC+1]} & \textbf{Duration} & \textbf{\# Initial Positions} & \textbf{\# Long ($\sim20\,$s) Trajectories / All} & \textbf{Mean / Median Speed [m/s]} \\
        \hline \hline
        \emph{LargeView} & December 8, 2022 20:13 & 20 s & 4081 & 114 / 277 & 0.44 / 0.36 \\
        \hline\hline
    \end{tabular}
    \caption{Basic statistics for the \emph{LargeView} dataset~\cite{data_madras_LargeViewTrajectories} inspected over the whole square.}
    \label{tab:CCTVstats}
\end{table}

\subsection{Complex flow in a region of high density near the center of the square}

Let us delve deeper into the pedestrian flows by examining the central zone, highlighted by the pink rounded square in  Fig.~\ref{fig:pedestrian_outflow_results}. This area, measuring $15$ meters by $25$ meters, exhibited complex patterns and high densities during two distinct periodic egresses. We tracked all pedestrians within this zone semi-manually, limited by the video resolution and occasional occlusions (see Sec.~\ref{sec:methods} for details). The extracted trajectories can be used to construct and animate a `digital twin' of the crowd. This animation\cite{streamlit} does not account for the heterogeneous sizes of pedestrians, representing all agents as standard adults, nor does it consider their social relationships. At first glance, there is a noticeable overall flow towards the West (the top of the image). However, the flow pattern is non-uniform, featuring counter-flows, individuals squeezing through the crowd, and others moving slowly.

\begin{figure}
    \centering
    \includegraphics[width=0.45\textwidth]{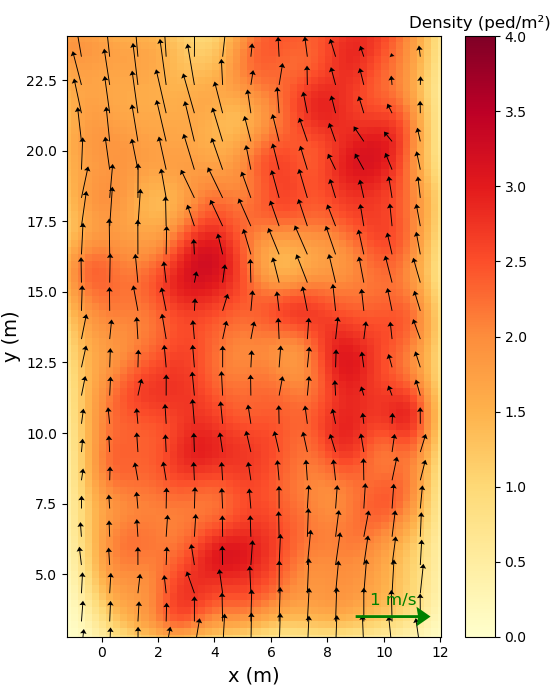}
    \includegraphics[width=0.45\textwidth]{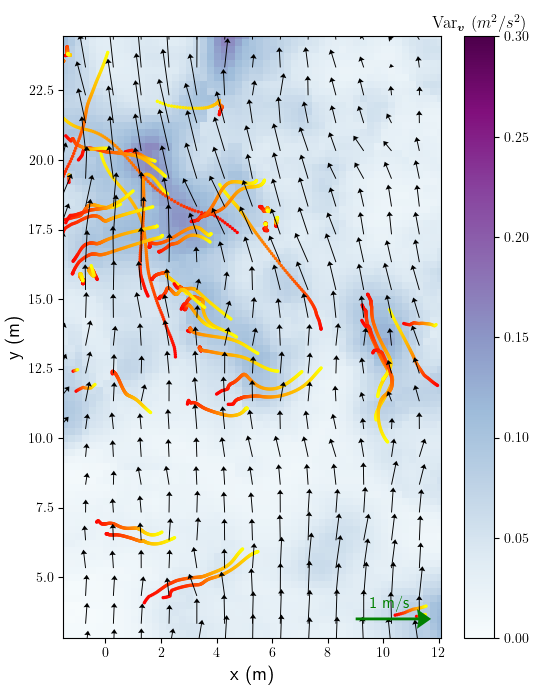}
    \caption{Continuous description of the complex flow at the center of the \emph{Place des Terreaux} at 20:15, recorded with a \emph{LargeView} camera (same orientation as Fig.~\ref{fig:PlTerreaux_full}). \textbf{Left panel:} Local density field averaged over the time window of 10 seconds in our local coordinate system. \textbf{Right panel:} Variance field $\mathrm{Var}_{\mathbf{v}}(\mathbf{r})$. The displayed trajectories (coloured from red to yellow as time moves on) are those of `counter-walking' pedestrians, i.e., those who significantly deviate from the continuous velocity field by  $\mathrm{Var}_{\mathbf{v}}^{i} \geqslant 0.12~ \mathrm{m^2/s^2}$. The arrows represent the continuous velocity field over the same time window of 10~s. All fields have been smoothed with a characteristic lengthscale $\xi =0.75\,\mathrm{m}$.}
    \label{fig:CCTV1}
\end{figure}

\begin{figure}
    \centering
    \includegraphics[width=0.45\textwidth]{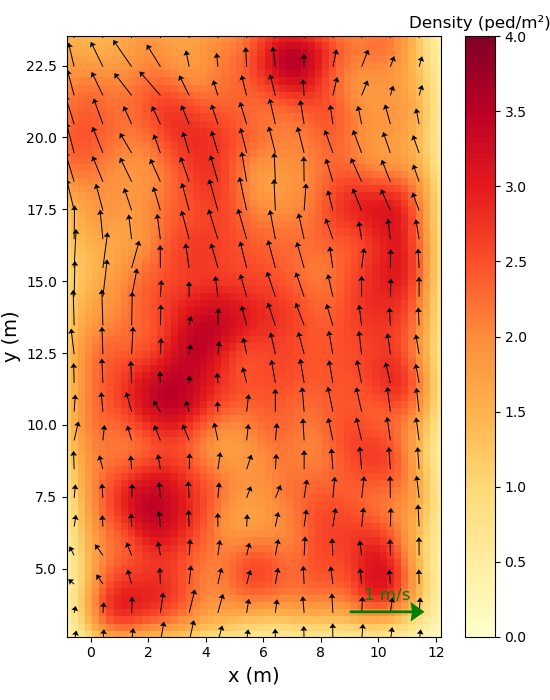}
    \includegraphics[width=0.45\textwidth]{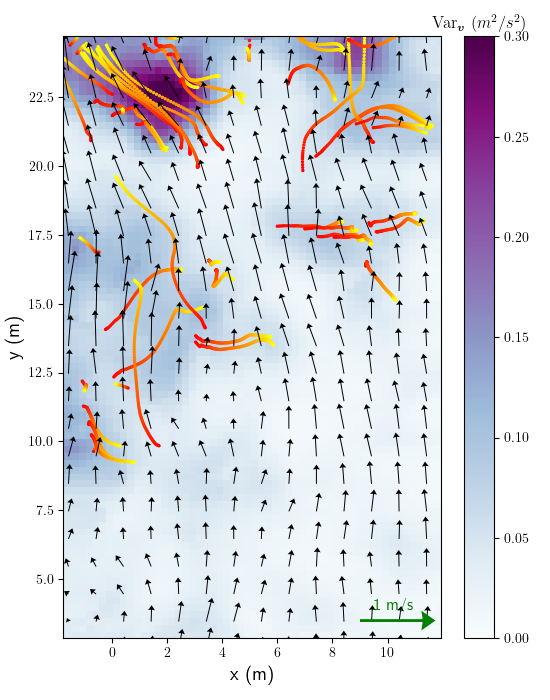}
    \caption{Continuous description of the complex flow at the center of the \emph{Place des Terreaux} at 21:05, recorded with a \emph{LargeView} camera. Refer to Fig.~\ref{fig:CCTV1} for the rest of the caption.}
    \label{fig:CCTV2}
\end{figure}

 The complex features that obscure the general characteristics can be simplified by coarse-graining trajectories into smooth density and velocity fields, as illustrated in the left panels of Fig.~\ref{fig:CCTV1} and Fig.~\ref{fig:CCTV2}. These smooth fields reveal noticeable density heterogeneities, but the flow pattern is more streamlined: most velocity vectors are aligned, directed towards the top of the figure, and rarely exceed half a meter per second. A slight tendency to navigate around densely populated areas is still observable. These coarse-grained fields can be interpreted as the underlying base flow. On top of this base flow, the variability of trajectories can be reintroduced by calculating local velocity variances, as shown in the right panels of Fig.~\ref{fig:CCTV1} and Fig.~\ref{fig:CCTV2}, highlighting trajectories that significantly deviate from the base flow (see Sec.~\ref{sec:methods}). This distinction between a smooth, streamlined base flow and counter-walking agents may be beneficial from a modelling perspective, allowing for advancements beyond the homogeneous flows predicted by macroscopic models.

The fundamental diagram relating the speed $\|\mathbf{v}\|(\mathbf{r},t)$ to the density $\rho(\mathbf{r},t)$ (with time binned into intervals of 0.5 seconds) is presented in Fig.~\ref{fig:PlTerreaux_FD}. It is essentially similar to that obtained for a sample of pedestrians across the entire square; the maximum speed observed at a given density exhibits a downward trend with increasing density, but all speeds below this upper bound are represented.

\begin{table}[!ht]
    \begin{tabular}{ p{2cm} p{1.7cm} p{1.7cm} p{2cm} p{2cm} p{1.8cm} p{3cm} }
        \hline 
        \textbf{File}  & \textbf{Start} {[}UTC+1{]}  & \textbf{Duration} & \textbf{\# trajectories}  & \textbf{Mean density} {[}ped/m$^{2}${]} & \textbf{Mean / median speed} {[}m/s{]}  & \textbf{Median / Mean trajectory duration} {[}s{]}\tabularnewline
        \hline \hline
        \emph{LargeView} Zoom\_A& December 8, 2022 20:16 & 20~s & 740 & 1.92 & 0.30 / 0.29 & 17.50 / 17.98\tabularnewline
        \emph{LargeView} Zoom\_O & December 9, 2022 21:05 & 45~s & 726 & 1.85 & 0.29 / 0.27 & 33.21 / 41.7\tabularnewline
        \hline \hline
    \end{tabular}
    \caption{Basic statistics for the exhaustive trajectory dataset~\cite{data_madras_LargeViewTrajectories} extracted from an area of interest in the \emph{LargeView} videos.}
    \label{tab:StatCCTVzoom} 
\end{table}

\subsection{Unidirectional and bidirectional flow at diverse densities along the northern building}

\begin{figure}[!ht]
    \centering
    \includegraphics[width=.95\linewidth]{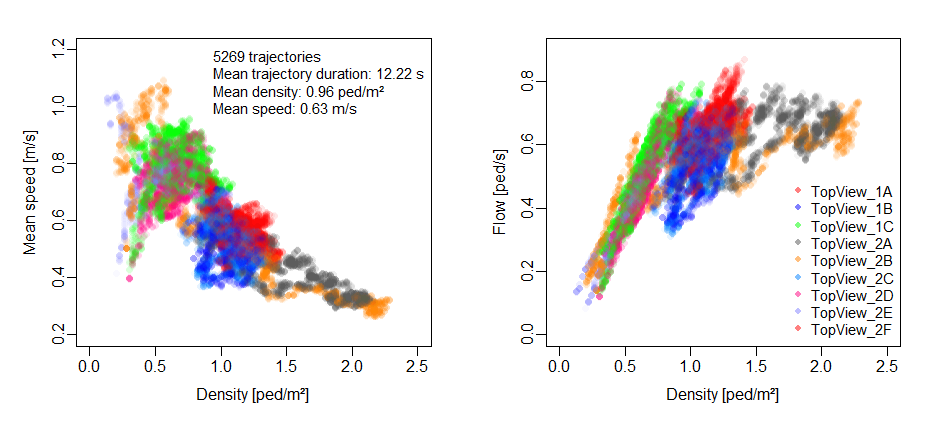}
    \caption{Fundamental diagram obtained by averaging the speed of the trajectories and the global density level in the scene over one-second time steps, for the \emph{TopView} datasets. }
    \label{fig:AirBB_FD}
\end{figure}

The flow pattern along the northern building of \emph{Place des Terreaux} (see Fig.~\ref{fig:outflow_extraction})  looks less complex compared to the preceding central region. In particular, there is a primary East-West direction. Besides, for these trajectories, the resolution and orientation of the camera (\emph{TopView} camera 2) afford higher-quality microscopic data. Notwithstanding this apparent relative simplicity, we will see that the crowd flows depart from their idealizations as unidirectional and bidirectional flows in controlled experiments, but to different extents. 
First, the fundamental diagram is presented in Fig.~\ref{fig:AirBB_FD}, which averages the speed of the trajectories and the \emph{global} density level in one-second intervals. This relationship displays much less scatter than the measurements at the center of \emph{Place des Terreaux} (see Fig.~\ref{fig:PlTerreaux_FD}). Furthermore, Table~\ref{tab:StatAirBB} provides basic statistics for each of the nine trajectory datasets, with three sequences selected explicitly for detailed analysis.

\begin{table}[!ht]
  \centering
    \begin{tabular}{|p{2cm}|p{1.7cm}|p{1.7cm}|p{2cm}|p{2cm}|p{1.8cm}|p{3cm}|}
      \hline
      \textbf{File} & \textbf{Start} [UTC+1] & \textbf{End} [UTC+1]& \textbf{\# trajectories} 
      & \textbf{Mean density} [ped/m$^2$]& \textbf{Mean speed} [m/s] & \textbf{Median / Mean trajectory duration} [s]\\ \hline \hline
      \emph{TopView}\_1A              & 22:40:45   &22:44:15     & 965&     1.13 &
0.57 &
       13.23 / 12.27             \\ \hline  
      \emph{TopView}\_1B            & 22:55:06    &22:57:46      & 685    &1.07&
0.52 &   11.76 / 12.64                 \\ \hline  
      \emph{TopView}\_1C                & 23:10:33   &23:13:58        & 673 & 0.65 & 0.78 & 9.2 / 9.03                          \\ \hline \hline
     \emph{TopView}\_2A                    & 21:26:27   &21:29:07    & 711 & 1.58 & 0.41&
 19.1 / 18.11              \\ \hline 
    \emph{TopView}\_2B                   & 21:40:39    &21:43:25      & 612 & 1.11 & 0.58&
  11.17 / 13.18             \\ \hline  
    \emph{TopView}\_2C                    & 22:55:18    &22:58:20      & 693& 1.06 & 0.54 &
 11.96 / 12.94                     \\ \hline  
    \emph{TopView}\_2D                    & 23:10:16    &23:12:57      & 529  & 0.64 & 0.75 & 8.83 / 8.96                \\ \hline 
    \emph{TopView}\_2E                    & 23:24:59    &23:26:11     & 218 & 0.57 & 0.75 & 8.83 / 8.63                       \\ \hline 
     \emph{TopView}\_2F                   & 23:54:59    &23:56:30     & 183  & 0.37 & 0.95 & 
6.8 / 6.92                 \\ \hline   \hline
    \multicolumn{3}{|r|}{Total}&5269 & 0.96 & 0.63 & 10.63 / 12.22\\
    \hline
    
    \end{tabular}
    \caption{Basic statistics for the exhaustive trajectory datasets~\cite{data_madras_TopViewTrajectories} extracted from \emph{TopView} videos.}
    \label{tab:StatAirBB}
  \end{table}


\paragraph{Unidirectional pedestrian flow (\emph{TopView}\_2B).}

\newcommand{\TrajectoryProfilePlot}[3]{
\begin{figure}[!ht]
\centering
\begin{minipage}[c]{.64\textwidth}
\centering
\includegraphics[width=\textwidth]{trajectories_#1.png}
\end{minipage}\hspace{.7cm}\begin{minipage}[c]{.27\textwidth}
\centering\footnotesize\vspace{-10mm}
Density profile$\qquad$\\[.5mm]
\includegraphics[width=\textwidth]{density_profile_#1.pdf}\\[2mm]
Velocity profile$\qquad$\\[-.25mm]
\includegraphics[width=\textwidth]{speed_profile_#1.pdf}\\
\end{minipage}\vspace{-2mm}
\caption{#2}
\label{#3}
\end{figure}
}

\newcommand{\TimeSeries}[3]{
\begin{figure}[!ht]
\centering
\includegraphics[width=.3\textwidth]{NT_#1.pdf}\hfill
\includegraphics[width=.3\textwidth]{density_#1.pdf}\hfill
\includegraphics[width=.3\textwidth]{speed_#1.pdf}
\caption{#2}
\label{#3}
\end{figure}
}

\newcommand{\Screenshot}[3]{
\begin{figure}[!ht]
\centering\medskip\small
\begin{minipage}[c]{.24\textwidth}
\centering
$t=0$\\[1mm]\includegraphics[width=\textwidth]{snapshot0_#1.png}\\[1mm]
$t=75$ s\\[1mm]\includegraphics[width=\textwidth]{snapshot75_#1.png}
\end{minipage}\hspace{7mm}\begin{minipage}[c]{.24\textwidth}
\centering
$t=25$ s\\[1mm]\includegraphics[width=\textwidth]{snapshot25_#1.png}\\[1mm]
$t=100$ s\\[1mm]\includegraphics[width=\textwidth]{snapshot100_#1.png}
\end{minipage}\hspace{7mm}\begin{minipage}[c]{.24\textwidth}
\centering
$t=50$ s\\[1mm]\includegraphics[width=\textwidth]{snapshot50_#1.png}\\[1mm]
$t=125$ s\\[1mm]\includegraphics[width=\textwidth]{snapshot125_#1.png}
\end{minipage}\smallskip
\caption{#2}
\label{#3}
\end{figure}
}

In the \emph{TopView}\_2B video recording, unidirectional flow prevails. Among the $612$ collected trajectories, $496$ pedestrians move from right to left, $60$ from left to right, $46$ from bottom to top, and $10$ from top to bottom. In the left panel of Fig.~\ref{fig:trajFekda6}, trajectories are colour-coded based on their entry and exit points: green for entry on the left and exit on the right, grey for entry on the right and exit on the left, red for entry from the bottom and exit at the top, and blue for entry from the top and exit at the bottom. Despite the prevailing unidirectionality, it is noteworthy that the streamlines are not strictly parallel in this essentially unconstrained geometry, even when focusing solely on the grey trajectories. Examining the density and speed profiles, computed using Gaussian-kernel filters (see Eq.~\eqref{eq:GaussianKernel} and the documentation in \cite{streamlit} and \cite{pedpy}), Fig.~\ref{fig:trajFekda6} (right panel) reveals that they are relatively uniform, except for a few pedestrians standing in the upper left of the scene, which results in density peaks.

Finally, Fig.~\ref{fig:seriesFekda6} presents a time series of various global indicators. The cumulative inflows from the four directions confirm the dominance of pedestrians moving to the right. Interestingly, there is a noticeable evolution of density over time. From $t=30\,\mathrm{s}$ onward, the density steadily increases from $0.4$ to $2.2$ ped/m$^2$, as observed in the snapshots of Fig.~\ref{fig:snapshotFekda6}. The average speed in the area mirrors this trend, decreasing steadily from about 1 to 0.3 m/s. These trends align with expectations based on existing literature on unidirectional flow in controlled experiments. The reduced average speed at low density can be ascribed to static individuals and the presence of social groups, which are known to reduce walking speed\cite{nicolas2023social}.

\TrajectoryProfilePlot
{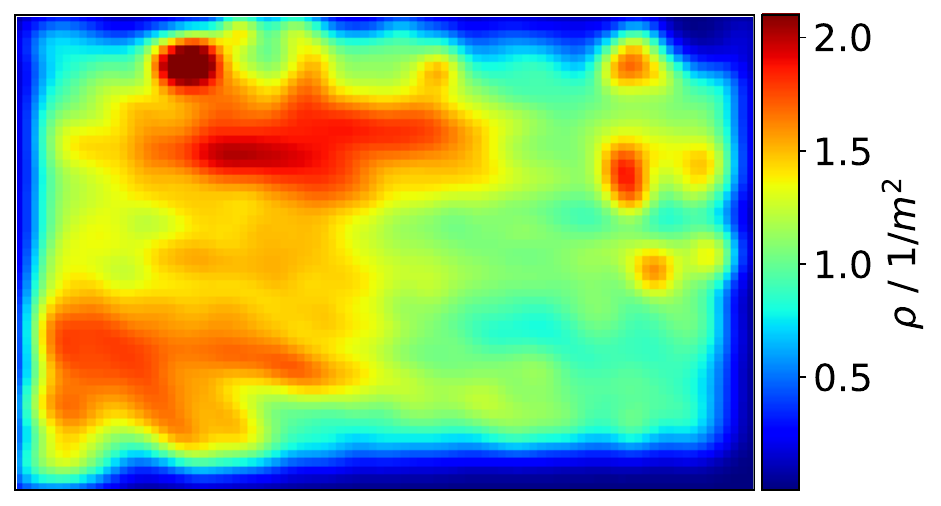}
{Trajectories \textbf{(left panel)} with density and speed profiles \textbf{(right panels)} for the \emph{TopView}\_2B video recording showing predominantly unidirectional pedestrian dynamics.}
{fig:trajFekda6}

\TimeSeries
{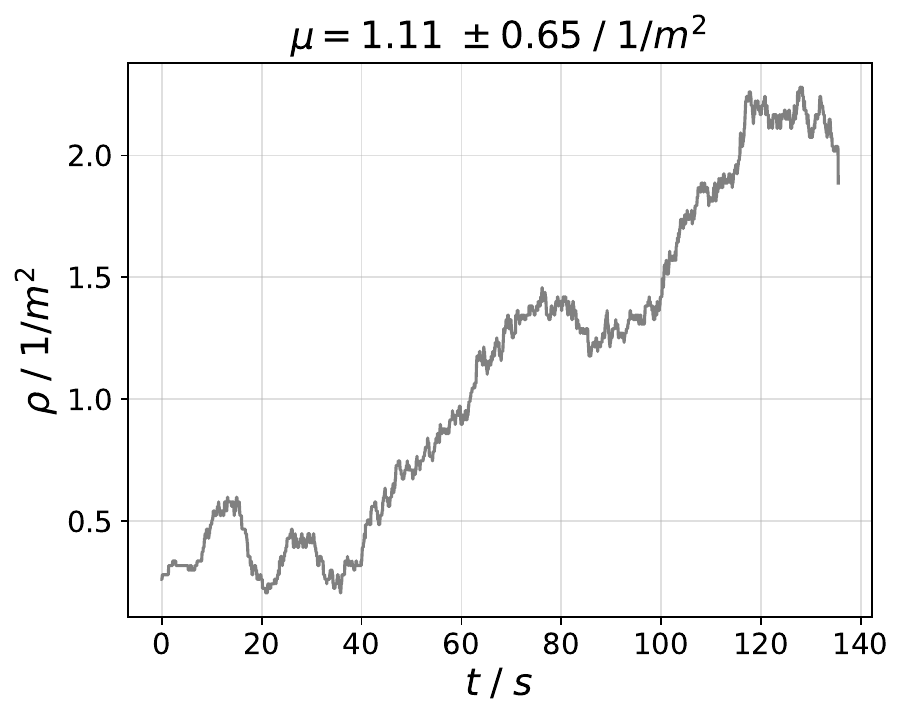}
{Cumulative flow, density and mean-speed time series for the \emph{TopView}\_2B video recording. The density increases over time for this video recording while the speed decreases.}
{fig:seriesFekda6}

\Screenshot
{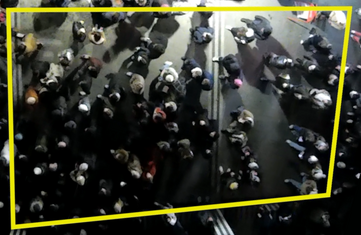}
{Snapshots of the \emph{TopView}\_2B video recording at different times.}
{fig:snapshotFekda6}

\paragraph{Unidirectional flow with standing pedestrians as obstacles (\emph{TopView}\_2C).}

The sequence \emph{TopView}\_2C, in the same zone, features an additional perturbation: a group of pedestrians standing in the upper left part of the scene. Out of the 693 trajectories, 603 walk to the left, 68 to the right, 15 up and seven down; the flow is thus predominantly unidirectional again. However, the static group has a conspicuous effect on the trajectories, forcing other pedestrians to go around it and causing congestion in the dynamics (see Fig.~\ref{fig:trajFekda10}, left panel). The density and speed profiles show two congested queues with reduced speed among the standing group (see Fig.~\ref{fig:trajFekda10}, right panels). The state is stationary in time, with a global density fluctuating between 0.8 and 1.3 ped/m$^2$ and a mean speed between 0.4 and 0.7~m/s (see Fig.~\ref{fig:seriesFekda10}). This is confirmed by the snapshots in Fig.~\ref{fig:snapshotFekda10}, which show similar crowding situations.

\TrajectoryProfilePlot{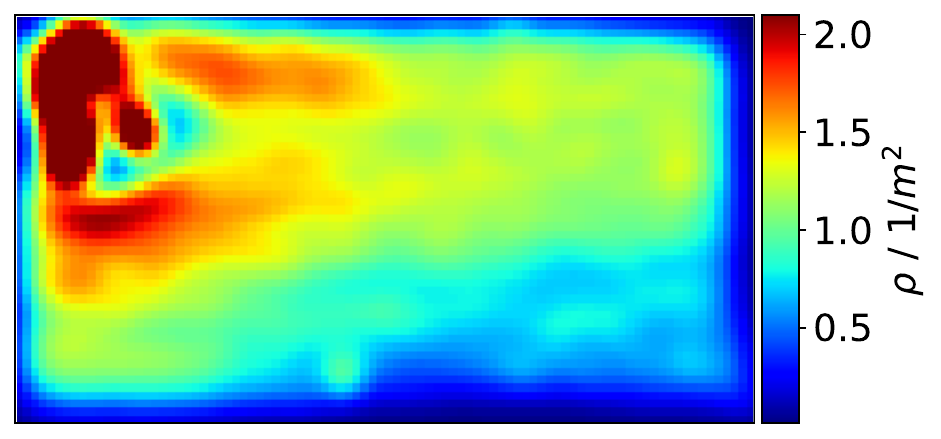}{Trajectories \textbf{(left panel)} with density and speed profiles \textbf{(right panels)} for the \emph{TopView}\_2C video recording. The pedestrian dynamics are predominantly unidirectional, while the scene includes standing pedestrians in the upper left corner, initiating avoidance behaviour and queuing.}
{fig:trajFekda10}

\TimeSeries
{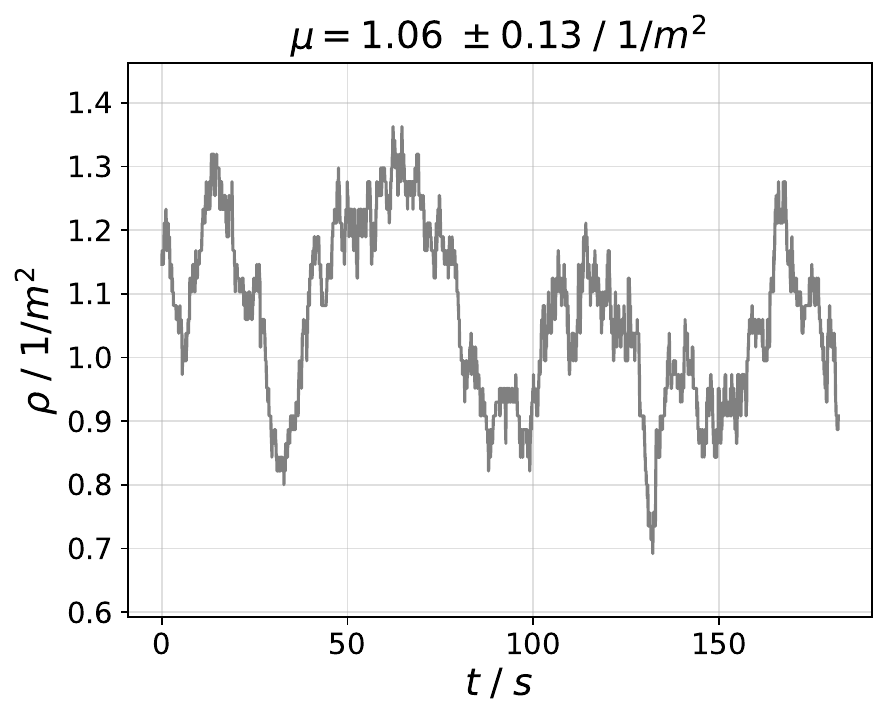}
{Cumulative flow, density and mean-speed time series for the \emph{TopView}\_2C video recording. The situation is relatively stationary in time.}
{fig:seriesFekda10}

\Screenshot
{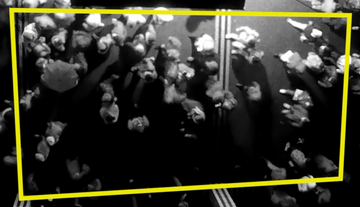}
{Snapshots of the \emph{TopView}\_2C video recording at different time instants.}
{fig:snapshotFekda10}

\paragraph{Unbalanced birectional flow (\emph{TopView}\_2D).}

The last recording that we analyze, Topview\_2D, contains 529 pedestrian trajectories, including 396 trajectories to the left, 109 to the right, 13 upward and 13 downward (see Figure~\ref{fig:trajFekda11}, left panel). The counter-walking pedestrians can no longer be neglected. They generate a substantial counterflow with lane formation by direction, separated by a group of standing pedestrians in the center right of the scene. Accordingly, we are dealing with an unbalanced (75\%:21\%) bidirectional flow.
The density and speed profiles are relatively homogeneous, although the flow to the left is slightly more congested (see Figure~\ref{fig:trajFekda11}, right panels). 
Again, the state is relatively stationary in time, with a global density fluctuating in between 0.4 and 1 ped/m$^2$ and a mean speed between 0.5 and 0.9~m/s (see Figure~\ref{fig:seriesFekda11}). 
The snapshots show that the crowd is sparser than in the previously presented video recordings (see Figure~\ref{fig:snapshotFekda11}).

\TrajectoryProfilePlot{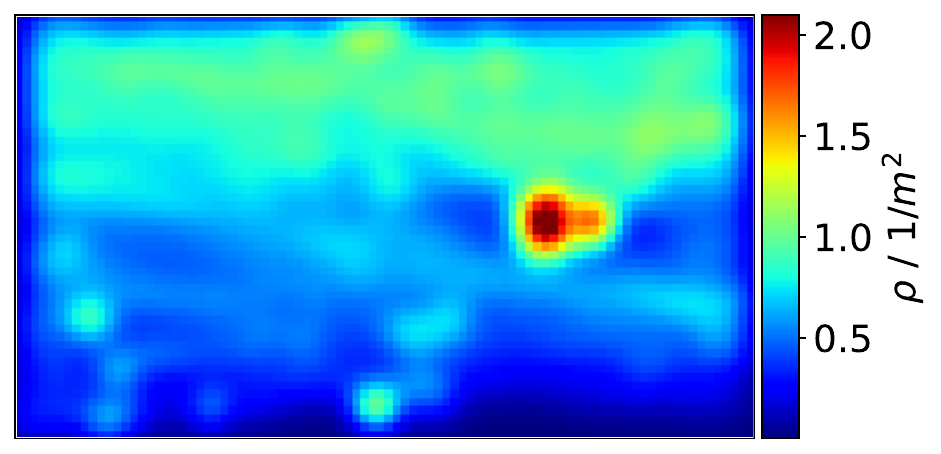}{Trajectories \textbf{(left panel)} with density and speed profiles \textbf{(right panels)} for the \emph{TopView}\_2D video recording (counterflow pedestrian dynamics with lane formation).}
{fig:trajFekda11}

\TimeSeries
{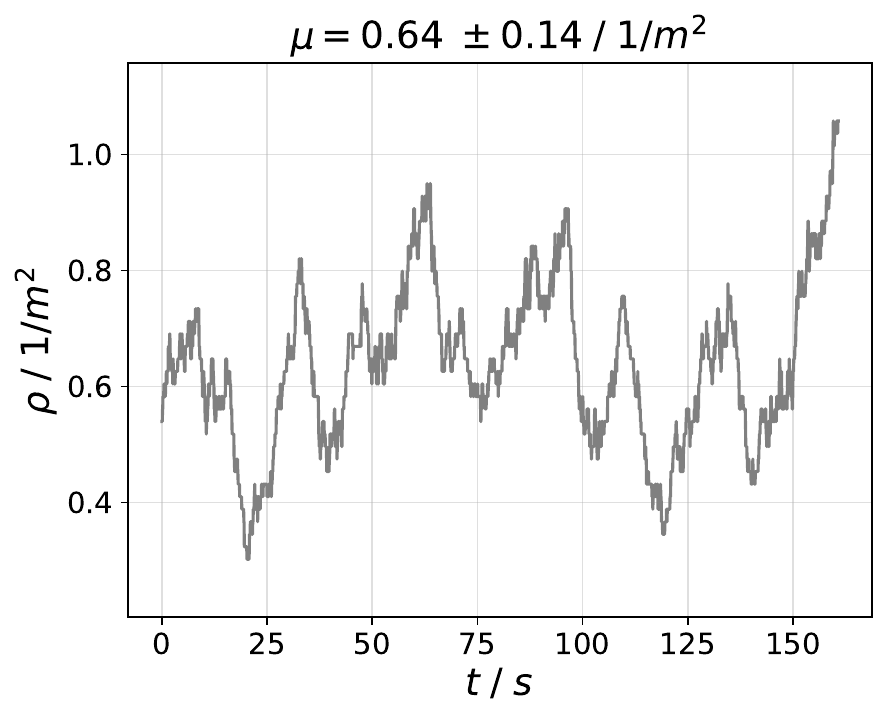}
{Cumulative flow, density and mean-speed time-series for the \emph{TopView}\_2D video recording. The situation is relatively stationary in time.}
{fig:seriesFekda11}

\Screenshot
{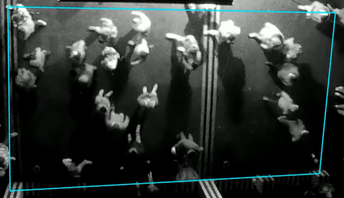}
{Snapshots of the \emph{TopView}\_2D video recording at different times.}
{fig:snapshotFekda11}

\subsection{Identification of singular qualitative phenomena}

This section highlights key features passively observed during the real-world Festival of Lights, a complex scenario that significantly diverges from controlled settings. 
Many of these features are due to the multidirectionality of the flow or, more generally, the fact that the pedestrians composing the crowd have different goals. These features notably include:
\begin{itemize}
    \item temporarily static groups of people around which passing pedestrians are forced to circumnavigate
     \item marked spatial variations of the density
    \item the serpentine motion of groups of people following each other through the crowd, thus moving like snakes (dubbed `crossing channels' in a controlled experimental study\cite{wang2023exploring} exploring crowd crossing)
    \item various non-standard pedestrians: people pushing a pushchair, pulling a piece of luggage, etc.
    \item a complex geometry of the premises that cannot strictly be reduced to two dimensions
    \item ambulances crossing the crowd
\end{itemize}

Depending on their prevalence and impact on pedestrian flow, these effects may need to be incorporated into models to achieve fully practical applications. We categorize these distinctive features into three groups: \textbf{(i) non-standard geometry of the premises}, \textbf{(ii) diversity of goals and speed preferences}, \textbf{(iii) heterogeneity of the crowd composition}. Additionally, we identify the video recordings and specific times when these features can be observed.

\subsubsection{Non-standard geometry of the premises}

In contrast to the common reliance on a binary geometry, which sets a binary distinction between accessible and inaccessible spaces, the square under study exhibits regions of varying attractiveness. Notably, the vicinity of the walls where shows are projected is visibly less appealing to the crowd. Some modeling approaches have been proposed in the literature to capture this heterogeneity\cite{helbing1997modelling,echeverria2023body}. Additionally, the geometry is neither composed of straight borders nor fully two-dimensional. Knee-high bollards and waist-high steel crowd barriers restrict movement (and are thus associated with lower local density, as shown in Fig.~\ref{fig:PlTerreaux_full}) but can overlap with pedestrians in three dimensions. The coordinates of these partial obstacles are provided in Dataset~\cite{data_madras_Geometry}.

\subsubsection{Diversity of goals and speed preferences}
Previously, we already highlighted the complexity of the flow patterns in some sequences and the multidirectionality of the flow. Here, we focus on the effects and consequences of the diversity of intentions among spectators, notably their diverse goals and speed preferences.

\paragraph{Static groups of people}

First and foremost, numerous temporarily static groups of people can be observed, often forcing passing pedestrians to navigate around them. These groups, typically consisting of $2$ to $8$ individuals, were found throughout the area of interest in the \emph{TopView} recordings. Two distinct scenarios can be identified: (i) a group moves, stops, and resumes motion; (ii) a group remains stationary for the entire footage duration. The first scenario is particularly intriguing, as it allows us to study the effects of people stopping, standing, and resuming motion over time. Table~\ref{tab:quali:groups} summarizes our observations of such static groups, excluding those who stop for less than two seconds. Although these groups are hardly included in controlled experiments of different flow types, they disrupt the base flow, significantly impacting the dynamics. Unlike classical obstacles, these groups are more complex because they are transient, appear and disappear, and fluctuate in size and shape over time. This variability can result from the addition of new members (see Table~\ref{tab:quali:groups}, \emph{TopView}\_2A file) or specific behaviours of group members, such as heckling (see Table~\ref{tab:quali:groups}, \emph{TopView}\_1B file). Additionally, the splitting of a group of standing people by walking pedestrians has also been witnessed (see Table~\ref{tab:quali:groups}, \emph{TopView}\_1C file).

\begin{table}
    \centering
    \begin{tabular}{ c c c p{10cm} }
        \hline
        \textbf{File}  & \textbf{Start} & \textbf{End} & \textbf{Description} \\ \hline\hline

        \emph{TopView}\_1B & 0:00 & 2:40 (end) & group of 7 standing on the right side. People are sometimes heckling or fighting, which makes the shape evolve (around 0:45). \\
        & & & Between 0:53 and 1:05, interactions between the standing group and a moving group of similar size. \\ \hline             
        \emph{TopView}\_1C & 0:00 & 0:56 & group of 3 people (2 adults and one child) standing in the middle. \\ \hline     
        \emph{TopView}\_1C & 0:25 & 0:50 & group of 3 people stopping, standing and restarting moving in the bottom right side. \\ \hline     
        \emph{TopView}\_1C & 0:58 & 1:30 & group of 2-standing people stopping, standing and resuming motion on the top side. This group is split by people passing through.\\ \hline     
        \emph{TopView}\_2A & 0:00 & 2:41 (end) & group of standing people on the left. The group size evolves from 2 to 8 people. \\ \hline     
        \emph{TopView}\_2B & 0:00 & 0:22 & group of 2 standing people on the left\\ \hline     
        \emph{TopView}\_2B & 0:45 & 1:12 & a group of 2 walking people stops and stands on the left \\ \hline        
        \emph{TopView}\_2C & 0:10 & 1:14 & a group of 2 walking people stops and stands on the left \\ \hline
        \emph{TopView}\_2E & 0:00 & 0:35 & 2 standing groups on the top (with limited impact on flow) \\ \hline
        \emph{TopView}\_2F & 0:00 & 0:53 & standing group of 3 people on the left border \\ \hline\hline
    \end{tabular}
    \caption{Situations of (temporarily) static groups observed in Dataset~\cite{data_madras_TopViewTrajectories}. }
    \label{tab:quali:groups}
\end{table}

\paragraph{Running pedestrians}
Conversely, we noticed that some people were running in the instances listed in Table~\ref{tab:my_label3}. 

\begin{table}[!ht]
    \centering
    \begin{tabular}{ c c c p{10cm} }
        \hline        
        \textbf{File}  & \textbf{Start} & \textbf{End} & \textbf{Nature} \\ \hline\hline
        \emph{TopView}\_1A & 2:49 & 2:53 & 2 people accelerating to reach a speed higher than the main flow  from middle to right side \\ 
        \emph{TopView}\_2B & 0:09 & 0:14 & 3 people running from left to right \\ \hline\hline
    \end{tabular}
    \caption{Instances of running pedestrians observed in Dataset~\cite{data_madras_TopViewTrajectories}. }
    \label{tab:my_label3}
\end{table}

\subsubsection{Marked density heterogeneities}

Giant density fluctuations far exceeding the fluctuations expected in a physical system at equilibrium are common in active matter assemblies  \cite{dey2012spatial,manning2023essay}. Here, marked density heterogeneities are conspicuous for the specific case of pedestrian assemblies. Some depleted regions (voids) are found not far from the high-density areas made of tightly packed groups in the same recording (Table~\ref{tab:my_label1}).

\begin{table}[!ht]
    \centering
    \begin{tabular}{ c c c p{10cm} }
        \hline      
        \textbf{File}  & \textbf{Start} & \textbf{End} & \textbf{Nature} \\ \hline\hline
        \emph{TopView}\_1A & 1:09 & 1:18  & gap, high-density in the left \\ 
        \emph{TopView}\_1B & 0:38 & 2:44  & high-density in the top left (due to standing people)\\ 
        \emph{TopView}\_2A & 0:02 & 0:27  & gap, high-density in the top (due to standing people)\\
        \emph{TopView}\_2C & 0:00 & 3:02  & high-density in the top left (due to standing people)\\ \hline\hline
    \end{tabular}
    \caption{Voids and density heterogeneities observed in Dataset~\cite{data_madras_TopViewTrajectories}.}
    \label{tab:my_label1}
\end{table}

\paragraph{Lines of people worming their way through the crowd (`serpentine' groups)}

In dense regions, we have often observed people worming their way through a static or counter-moving crowd and following each other, thus forming linear, snake-like structures (Table~\ref{tab:my_label2}). People follow each other along these linear structures, dubbed serpentine groups, most probably due to the depleted channels opened in the wake of their predecessors \cite{nicolas2019mechanical} and their possible social relationships.

Similar self-organized structures have been observed in controlled experiments of people crossing static groups and dubbed `crossing channels\cite{wang2023exploring},' but overall, they have received much less attention than stable lanes in bidirectional flows or the stripes formed at the intersection of two flows. Indeed, their frequency in the empirical dataset seems to owe much to the multiple directions of pedestrians and the non-stationary character of the flow. We hypothesize that the observed transient `snakes' could turn into stable lanes in stationary conditions and with a limited number of directions, with distinct consequences on the flow properties.

\begin{table}[!ht]
    \centering
    \begin{tabular}{ c c c p{10cm} }
        \hline        
        \textbf{File}  & \textbf{Start} & \textbf{End} & \textbf{Description} \\ \hline  \hline
        \emph{TopView}\_1A & 0:35 & 0:47 & lane formation on both sides \\ \hline
        \emph{TopView}\_1A & 1:11 & 1:28 & mini-lanes: 3 people walk counter to the main flow \\ \hline         
        \emph{TopView}\_1A & 1:12 & 2:05 & serpentine group at the top \\ \hline 
        \emph{TopView}\_1B & 1:05 & 2:40 (end) & snake/lane formation from right to left side due to a standing group; quite high density.\\ \hline
                
        \emph{TopView}\_2A & 0:24 & 0:39 & a group of 6 is worming their way through a dense counter-moving crowd. \\ \hline
        
        \emph{TopView}\_2B & 0:27 & 0:40 & a group of 7 is worming their way through a crowd moving in the same direction. The group splits (going from bottom to right side). \\ \hline
        \emph{TopView}\_2B & 1:12 & 2:00 & a lane appears on the top, opposite the main flow. \\ \hline
    
        \emph{TopView}\_2C & 0:27 & 0:40 & a serpentine group of 2-3 people at moderately high density (from right to left) \\ \hline
        \emph{TopView}\_2C & 0:00 & 0:15 & a serpentine group of 9 at medium density (from right to left) \\ \hline\hline
    \end{tabular}
    \caption{Serpentine groups (people walking counter to the main flow and following each other in line)  observed in Dataset~\cite{data_madras_TopViewTrajectories}.}
    \label{tab:my_label2}
\end{table}

\subsubsection{Heterogeneity of the crowd composition}

\paragraph{Social groups}

Unlike the homogeneous crowds of individual agents traditionally considered by crowd modelers, the crowd at the Festival of Lights primarily consists of social groups. Some groups are quite large (see Fig.~\ref{fig:group_sizes}), even though they may be split in practice, and some are families with children. Naturally, this composition is expected to influence (at least) the microscopic dynamics at play.

\paragraph{Pushchairs and bikes}

 Moreover, not all pedestrians fit the standard image of a typical pedestrian. Some navigate through crowds of varying densities while pushing strollers, while others maneuver their bicycles  (Table~\ref{tab:my_label4}). Consequently, the shape of the agent to be modeled differs widely from that of a standard pedestrian. Additionally, the density tends to be higher in front of strollers than behind them, prompting pedestrians behind the `pushers' to frequently try to overtake them.

\begin{table}[!ht]
    \centering
    \begin{tabular}{ c c c c }
        \hline        
        \textbf{File}  & \textbf{Start} & \textbf{End} & \textbf{Nature} \\ \hline \hline  
        \emph{TopView}\_1B & 0:01 & 0:13 & 
        pedestrian with a bike \\ 
        \emph{TopView}\_1B & 1:33 & 1:48 & 
        pushchair \\ 
        \emph{TopView}\_1C & 3:05 & 3:16 & 
        pushchair \\       
        \emph{TopView}\_2A & 1:25 & 1:53 & 
        pushchair \\   
        \emph{TopView}\_2A & 2:33 & 2:41 (end) & 
        pushchair \\ 
        \emph{TopView}\_2B & 0:49 & 1:12 & 
        pushchair \\
        \emph{TopView}\_2E & 0:36 & 0:50 & pushchair \\ 
        \emph{TopView}\_2F & 0:59 & 1:06 & 
        pedestrian with a bike \\ \hline\hline
    \end{tabular}
    \caption{Non-standard pedestrians (pushchairs, bikes, etc.)  observed in Dataset~\cite{data_madras_TopViewTrajectories}.}
    \label{tab:my_label4}
\end{table}

\paragraph{Ambulances}
Finally, we observed instances where an ambulance needed to cross the crowd. In response, a channel opened in the crowd ahead of the vehicle to let it through (\url{https://www.youtube.com/watch?v=1zqpJRnAqsM}).

\section{Technical Validation}

\paragraph{Trajectory datasets for the \emph{TopView} recordings}

\begin{figure}[!ht]
    \centering\bigskip
    \includegraphics[height=5.5cm]{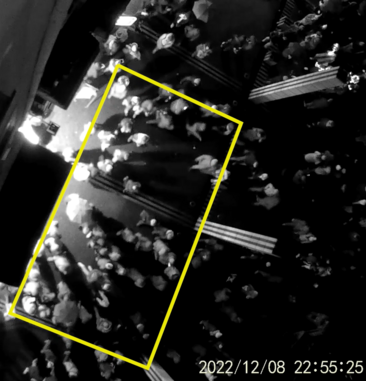}\hspace{2cm}
    \includegraphics[height=5.5cm]{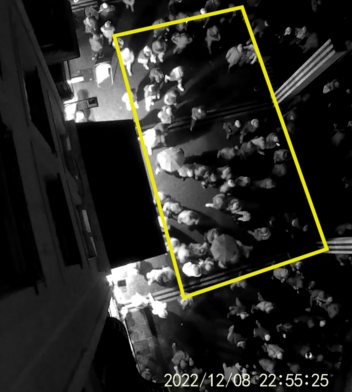}\medskip
    \caption{Point of view for the \emph{TopView}\_1B and \emph{TopView}\_2C camera videos, which simultaneous captured the same scene from different locations and angles.}
    \label{fig:PointOfViewAirBnB}
\end{figure}

\begin{figure}[!ht]
    \centering
    \includegraphics{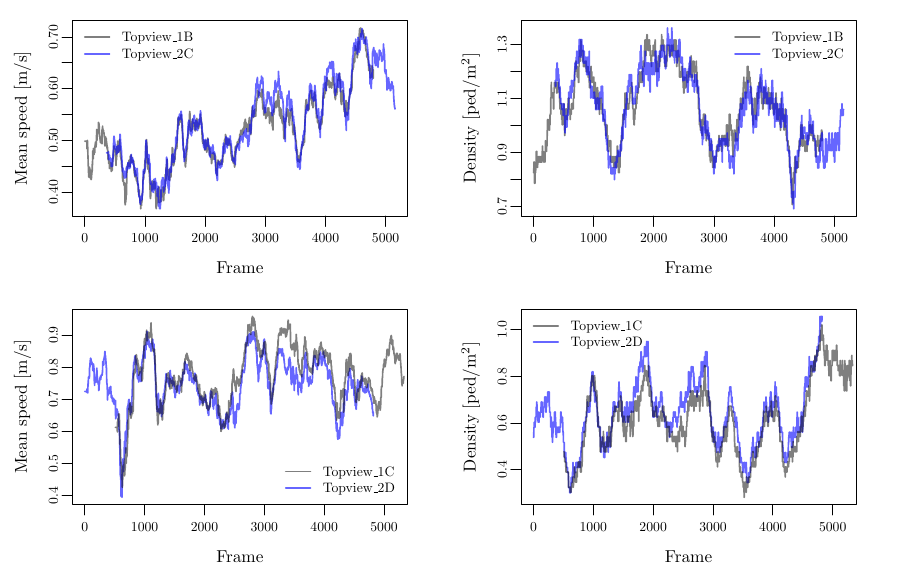}
    \caption{Superposition of the pedestrian mean speed \textbf{(left panels)} and density \textbf{(right panels)} time series for the \emph{TopView}\_1B and \emph{TopView}\_2C video recordings (upper panels) and for the \emph{TopView}\_1C and \emph{TopView}\_2D video recordings (lower panels), which capture the same scene from different locations and angles of view, and which partly overlap in time.}
    \label{fig:Verif}
\end{figure}

All trajectory datasets were visually inspected and manually corrected where necessary. In addition, comparing the overlapping time series of density and mean speed obtained from the \emph{TopView}\_1B and \emph{TopView}\_2C  video recordings, on the one hand, and the \emph{TopView}\_1C and \emph{TopView}\_2D videos, on the other hand, further validated the results (Fig.~\ref{fig:Verif}). The fields of view of these cameras are similar (see Fig.~\ref{fig:PointOfViewAirBnB}) and their recordings overlap in time (see Table~\ref{tab:StatAirBB}). We find a relatively small root-mean-square differences (from 5 to 10\%) between overlapping sequences, both for the mean speed and for the density (see Table~\ref{tab:RMSE_verif}). Nonetheless, a systematic bias is observed, in opposite directions for \emph{TopView}\_1B/\emph{TopView}\_2C and \emph{TopView}\_1C/\emph{TopView}\_2D. 
These biases may be explained by perspective effects and partly biased correction of optical and geometric distortions.

\begin{table}[!ht]
    \centering
    \begin{tabular}{c c c }
        \hline
        \textbf{File} & \textbf{RMSD Mean Speed} [$\mathrm{m/s}$] & \textbf{RMSD Density} [$\mathrm{ped/m}^2$] \\
        \hline\hline
        \emph{TopView}\_1B / \emph{TopView}\_2C & 0.02 & 0.05 \\ 
        \emph{TopView}\_1C / \emph{TopView}\_2D & 0.05 & 0.07 \\ 
        \hline\hline
    \end{tabular}
    \caption{Root mean square differences (RMSD) between the pedestrian mean speed and density time series for the \emph{TopView}\_1B and \emph{TopView}\_2C video recordings and the \emph{TopView}\_1C and \emph{TopView}\_2D video recordings, as shown in Fig.~\protect\ref{fig:Verif}.}
    \label{tab:RMSE_verif}
\end{table}

\paragraph{Trajectory datasets for the \emph{LargeView} recordings} As mentioned above, owing to the lower resolution of the \emph{LargeView} recordings, the quality of the extracted data is not quite as good. Occasionally, we may have failed to detect shorter individuals or swapped intersecting trajectories. Despite these challenges, two staff members (called `coders') independently extracted trajectories from different recordings and then analyzed each other's work. The coders largely agreed on the extracted data, although there were occasional disagreements or uncertainties regarding some data points. The common trends observed in the density and velocity fields, as shown in Figs.~\ref{fig:CCTV1}-\ref{fig:CCTV2}, which correspond to different days of recording, further support the robustness of the presented data.

To further validate the \emph{LargeView} trajectory dataset, possible detection or tracking errors were analyzed jointly by the two coders in a second stage. Errors were categorized as either major or minor. Major errors included omissions of clearly visible pedestrians and the creation of non-existent ones. Minor errors involved misidentifying or possibly swapping pedestrians as well as slightly inaccurate clicks on a pedestrian's head. The joint analysis led to the detection of 
10 major errors ($1.3\%$) out of $\sim 740$ trajectories and 6 minor errors in the  \emph{LargeView} Zoom\_O dataset, and 
3 major errors ($0.4\%$) out of $\sim 740$ trajectories and 5 minor errors in the \emph{LargeView} Zoom\_A dataset, over 10 seconds. (All these errors were corrected in the final dataset.)

In order to assess errors on the local densities, two subregions of rectangular shape were defined at distinct locations in time and space, each measuring $4\times 6$ m$^2$, and the two coders separately counted all people in these regions, including flickering appearances who were \emph{likely} to be people, even if this was not certain, in order to get an upper bound.
Their respective counts typically differed by less than 10\%, and exceeded the number of actually tracked pedestrians by 20\% to 30\% in the \emph{LargeView} Zoom\_O dataset and 9\% to 12\% in the \emph{LargeView} Zoom\_A dataset, depending on the location of the rectangle.
This leads to the conclusion that the local densities given by our dataset underestimate the actual densities by \emph{at most} by 9\% to 30\%. 

\paragraph{Mapping to real-world coordinates} To map the pedestrian positions in pixel coordinates to real-world coordinates, calibration using people standing at predefined positions was performed; the distances between the predefined positions were carefully measured on the ground. In the most distant part of the square, the calibration error on the real-world coordinates (but not the \emph{relative} positions) may reach a couple of meters. Then, after conversion, we successfully checked the compatibility of the crowd positions with the geometry of the premises obtained from \emph{Google Earth} data and our independent positioning of obstacles.

\paragraph{Surveys}
Six distinct staff members gave out oral surveys about origins, destinations, and group sizes. In addition to the collected statements, group sizes were also passively observed on the field.

\section{Usage Notes}


\begin{description}[nosep]
    \item[Online platform to visualise the data\cite{streamlit}:] We have released an open online platform that allows users to visualize and plot most of the collected data.
    
    \item[Survey results:] The survey results are provided in a CSV file,  with each line representing a respondent. A README file is included to help interpret the columns.
    
    \item[Geometry of the premises] The geometry of the square and the (temporary) obstacles are provided as CSV files using the WKT format for geometries. 
    
    \item[GPS traces and statistics of physical contacts:] Contact data are compiled in a CSV file,  detailing the duration, end time, and instances of contact. A README file is included to assist with column interpretation. GPS data for $10$ individuals, formatted in GPX, are linked with the contact data. Note that some individuals in the contact data did not provide GPS trajectories. 
    
    \item[Trajectory data:] Each file is identified by a number, which corresponds to the camera and recording location (see Fig.~\ref{fig:maps}), and a letter, which indicates the sequence (i.e., time). The videos from which the trajectories were extracted are also available. Trajectories are given in real-world coordinates at a frequency of $10$ Hz after all geometric corrections and interpolation have been applied, but without any smoothing or filtering. They are available in both local coordinates (suffix `loc'), used to plot most figures in this chapter, and absolute coordinates (EPSG:2154) for global location. The time reference is set to Friday 9 December 2022, at 8 pm, with time given in seconds before or after this reference.

    \item[Additional videos:] Many more videos were collected than we could analyze. We welcome contributions from volunteers to assist with the tracking efforts. We have made our tracking and camera calibration scripts available to support this. 
\end{description}

\section{Code availability}


Custom code was developed to generate, process, and analyze the datasets presented in this study. Additionally, a Streamlit application enables interactive data exploration and access under specific conditions. The source code is available on GitHub~\cite{streamlit} under the MIT license, allowing  noncommercial use, distribution, and modification with proper attribution. The repository includes:
\begin{itemize}[nosep]
    \item The scripts and notebooks used for data collection, preprocessing, analysis, and visualization.
    \item The Streamlit application code, along with instructions for installation and deployment \cite{streamlit}.
    \item A requirements.txt file specifying the versions of the software and libraries used to ensure compatibility. Essential software versions include PedPy 1.0.2~\cite{pedpy} as a backend for speed, density and flow calculations.
\end{itemize} 
The repository also includes comprehensive documentation that details the specific variables, parameters, and settings used to facilitate the reproduction of the results and analyses presented in this study. This encompasses but is not limited to, parameters for data filtering and the statistical analysis methods applied. While the code is publicly accessible, we encourage users to adhere to the terms of the license agreement, especially regarding non-commercial use and the requirement for proper attribution.




\section*{Acknowledgements} 

We acknowledge the precious help of Clément Albin, David Rodney, and Nicolas Verstaevel, as well as that of the technical municipal services of Lyon and organizers of the Fête des Lumières (Christophe Doucet, Thierry Courtot, Loïc Malacher, Antony Sabater).
We are also grateful to Maik Boltes and Alessandro Corbetta for their advice on the choice of hardware and to Tobias Schrödter for his support in using \emph{PeTrack}.
The authors acknowledge the Franco-German research project MADRAS, funded in France by the Agence Nationale de la Recherche (ANR, French National Research Agency), grant number ANR-20-CE92-0033, and in Germany by the Deutsche Forschungsgemeinschaft (DFG, German Research Foundation), grant number 446168800.

\section*{Author contributions statement}


Each team member played a significant role in various aspects of the project, including data collection, analysis, writing, and tool development. All authors reviewed the manuscript, participated in data acquisition, and ensured the accuracy of the collected data. Below is a summary of the contributions:
\begin{description}[nosep]
    \item[O.D.:] Organized data collection in Lyon, tracked \emph{LargeView} trajectories manually, collected and analyzed contact numbers, GPS tracks, and \emph{LargeView} trajectories, contributed to the Festival of Lights description and gathering methodology, calibrated \emph{LargeView} cameras and contributed to build an interactive app for data exploration.
    \item[H.D.:] Calibrated the \emph{TopView} cameras, extracted \emph{TopView} trajectories using \emph{PeTrack}, and contributed to the gathering methodology.
    \item[J.C.:] Selected time sequences and zones for analysis, extracted trajectories using \emph{PeTrack}, and contributed to the gathering methodology.
    \item[R.K.:] Participated in data acquisition, extracted \emph{TopView} trajectories using \emph{PeTrack} and contributed to the literature survey and gathering methodology.
    \item[M.C.:] Developed an interactive app for data exploration, extracted \emph{TopView} trajectories using \emph{PeTrack}, and contributed to the analysis of quantitative trajectory data.
    \item[B.G.:] Monitored pedestrian flow, contributed to the Festival of Lights description.
    \item[A.N.:] Organized data collection in Lyon, participated in data acquisition, manually collected and analyzed \emph{LargeView} trajectories, manually tracked heads in the square scene, and contributed to identifying statistics and qualitative phenomena.
    \item[A.T.:] Contributed to the literature survey, extracted \emph{TopView} trajectories using \emph{PeTrack}, quantitative data analysis, data extraction validation, and indicators/measures formulae development.
\end{description}

\section*{Competing interests}
The authors have interacted with the organizers of the Festival of Lights, but have not received any funding from them. They are hierarchically and financially independent from them and are not aware of any other competing interests.



\newcommand{\TimeSeriesScreenshot}[3]{
\begin{figure}[!ht]
\centering
\begin{minipage}[c]{.35\textwidth}
\centering
\includegraphics[width=\textwidth]{density_#1.pdf}\\[-2mm]
\includegraphics[width=\textwidth]{speed_#1.pdf}\\[-2mm]
\includegraphics[width=\textwidth]{NT_#1.pdf}
\end{minipage}\hfill\begin{minipage}[c]{.25\textwidth}
$t=0$\\[1mm]\includegraphics[width=\textwidth]{snapshot0_#1.png}\\[1mm]
$t=25$ s\\[1mm]\includegraphics[width=\textwidth]{snapshot25_#1.png}\\[1mm]
$t=50$ s\\[1mm]\includegraphics[width=\textwidth]{snapshot50_#1.png}\\[1mm]
$t=75$ s\\[1mm]\includegraphics[width=\textwidth]{snapshot75_#1.png}
\end{minipage}\hspace{10mm}\begin{minipage}[c]{.25\textwidth}
$t=100$ s\\[1mm]\includegraphics[width=\textwidth]{snapshot100_#1.png}\\[1mm]
$t=125$ s\\[1mm]\includegraphics[width=\textwidth]{snapshot125_#1.png}\\[1mm]
$t=150$ s\\[1mm]\includegraphics[width=\textwidth]{snapshot125_#1.png}\\[1mm]
$t=175$ s\\[1mm]\includegraphics[width=\textwidth]{snapshot125_#1.png}
\end{minipage}
\label{#3}
\caption{#2}
\end{figure}}


\begin{thebibliography}{}
\urlstyle{rm}
\expandafter\ifx\csname url\endcsname\relax
  \def\url#1{\texttt{#1}}\fi
\expandafter\ifx\csname urlprefix\endcsname\relax\def\urlprefix{URL }\fi
\expandafter\ifx\csname doiprefix\endcsname\relax\def\doiprefix{DOI: }\fi
\providecommand{\bibinfo}[2]{#2}
\providecommand{\eprint}[2][]{\url{#2}}

\end{thebibliography}


\begin{thebibliography}{10}
\urlstyle{rm}
\expandafter\ifx\csname url\endcsname\relax
  \def\url#1{\texttt{#1}}\fi
\expandafter\ifx\csname urlprefix\endcsname\relax\def\urlprefix{URL }\fi
\expandafter\ifx\csname doiprefix\endcsname\relax\def\doiprefix{DOI: }\fi
\providecommand{\bibinfo}[2]{#2}
\providecommand{\eprint}[2][]{\url{#2}}

\bibitem{helbing2007dynamics}
\bibinfo{author}{Helbing, D.}, \bibinfo{author}{Johansson, A.} \&
  \bibinfo{author}{Al-Abideen, H.~Z.}
\newblock \bibinfo{journal}{\bibinfo{title}{Dynamics of crowd disasters: An
  empirical study}}.
\newblock {\emph{\JournalTitle{Physical review E}}}
  \textbf{\bibinfo{volume}{75}}, \bibinfo{pages}{046109},
  \url{https://doi.org/10.1103/PhysRevE.75.046109} (\bibinfo{year}{2007}).

\bibitem{sieben2023inside}
\bibinfo{author}{Sieben, A.} \& \bibinfo{author}{Seyfried, A.}
\newblock \bibinfo{journal}{\bibinfo{title}{Inside a life-threatening crowd:
  Analysis of the love parade disaster from the perspective of eyewitnesses}}.
\newblock {\emph{\JournalTitle{Safety science}}}
  \textbf{\bibinfo{volume}{166}}, \bibinfo{pages}{106229},
  \url{https://doi.org/10.1016/j.ssci.2023.106229} (\bibinfo{year}{2023}).

\bibitem{feliciani2023trends}
\bibinfo{author}{Feliciani, C.}, \bibinfo{author}{Corbetta, A.},
  \bibinfo{author}{Haghani, M.} \& \bibinfo{author}{Nishinari, K.}
\newblock \bibinfo{journal}{\bibinfo{title}{Trends in crowd accidents based on
  an analysis of press reports}}.
\newblock {\emph{\JournalTitle{Safety science}}}
  \textbf{\bibinfo{volume}{164}}, \bibinfo{pages}{106174},
  \url{https://doi.org/10.1016/j.ssci.2023.106174} (\bibinfo{year}{2023}).

\bibitem{sharma2023global}
\bibinfo{author}{Sharma, A.} \emph{et~al.}
\newblock \bibinfo{journal}{\bibinfo{title}{Global mass gathering events and
  deaths due to crowd surge, stampedes, crush and physical injuries-lessons
  from the seoul halloween and other disasters}}.
\newblock {\emph{\JournalTitle{Travel medicine and infectious disease}}}
  \textbf{\bibinfo{volume}{52}},
  \url{https://doi.org/10.1016/j.tmaid.2022.102524} (\bibinfo{year}{2023}).

\bibitem{fruin1970designing}
\bibinfo{author}{Fruin, J.~J.}
\newblock \bibinfo{journal}{\bibinfo{title}{Designing for pedestrians: A
  level-of-service concept}}.
\newblock {\emph{\JournalTitle{Highway Research Record}}}
  \textbf{\bibinfo{volume}{355}}, \bibinfo{pages}{1--15},
  \url{https://onlinepubs.trb.org/Onlinepubs/hrr/1971/355/355-001.pdf}
  (\bibinfo{year}{1971}).
\newblock \bibinfo{note}{Presented at the 50th Annual Meeting}.

\bibitem{haghani2020empiricalI}
\bibinfo{author}{Haghani, M.}
\newblock \bibinfo{journal}{\bibinfo{title}{Empirical methods in pedestrian,
  crowd and evacuation dynamics: Part i. experimental methods and emerging
  topics}}.
\newblock {\emph{\JournalTitle{Safety science}}}
  \textbf{\bibinfo{volume}{129}}, \bibinfo{pages}{104743},
  \url{https://doi.org/10.1016/j.ssci.2020.104760} (\bibinfo{year}{2020}).

\bibitem{garcimartin2018redefining}
\bibinfo{author}{Garcimart{\'\i}n, A.} \emph{et~al.}
\newblock \bibinfo{journal}{\bibinfo{title}{Redefining the role of obstacles in
  pedestrian evacuation}}.
\newblock {\emph{\JournalTitle{New Journal of Physics}}}
  \textbf{\bibinfo{volume}{20}}, \bibinfo{pages}{123025},
  \url{https://doi.org/10.1088/1367-2630/aaf4ca} (\bibinfo{year}{2018}).

\bibitem{zhang2014comparison}
\bibinfo{author}{Zhang, J.} \& \bibinfo{author}{Seyfried, A.}
\newblock \bibinfo{journal}{\bibinfo{title}{Comparison of intersecting
  pedestrian flows based on experiments}}.
\newblock {\emph{\JournalTitle{Physica A: Statistical Mechanics and its
  Applications}}} \textbf{\bibinfo{volume}{405}}, \bibinfo{pages}{316--325},
  \url{https://doi.org/10.1016/j.physa.2014.03.004} (\bibinfo{year}{2014}).

\bibitem{cao2017fundamental}
\bibinfo{author}{Cao, S.}, \bibinfo{author}{Seyfried, A.},
  \bibinfo{author}{Zhang, J.}, \bibinfo{author}{Holl, S.} \&
  \bibinfo{author}{Song, W.}
\newblock \bibinfo{journal}{\bibinfo{title}{Fundamental diagrams for
  multidirectional pedestrian flows}}.
\newblock {\emph{\JournalTitle{Journal of Statistical Mechanics: Theory and
  Experiment}}} \textbf{\bibinfo{volume}{2017}}, \bibinfo{pages}{033404},
  \url{https://doi.org/10.1088/1742-5468/aa620d} (\bibinfo{year}{2017}).

\bibitem{haghani2020evacuation}
\bibinfo{author}{Haghani, M.}, \bibinfo{author}{Sarvi, M.} \&
  \bibinfo{author}{Shahhoseini, Z.}
\newblock \bibinfo{journal}{\bibinfo{title}{Evacuation behaviour of crowds
  under high and low levels of urgency: Experiments of reaction time, exit
  choice and exit-choice adaptation}}.
\newblock {\emph{\JournalTitle{Safety science}}}
  \textbf{\bibinfo{volume}{126}}, \bibinfo{pages}{104679},
  \url{https://doi.org/10.1016/j.ssci.2020.104679} (\bibinfo{year}{2020}).

\bibitem{pastor2015experimental}
\bibinfo{author}{Pastor, J.~M.} \emph{et~al.}
\newblock \bibinfo{journal}{\bibinfo{title}{Experimental proof of
  faster-is-slower in systems of frictional particles flowing through
  constrictions}}.
\newblock {\emph{\JournalTitle{Physical Review E}}}
  \textbf{\bibinfo{volume}{92}}, \bibinfo{pages}{062817},
  \url{https://doi.org/10.1103/PhysRevE.92.062817} (\bibinfo{year}{2015}).

\bibitem{murakami2021mutual}
\bibinfo{author}{Murakami, H.}, \bibinfo{author}{Feliciani, C.},
  \bibinfo{author}{Nishiyama, Y.} \& \bibinfo{author}{Nishinari, K.}
\newblock \bibinfo{journal}{\bibinfo{title}{Mutual anticipation can contribute
  to self-organization in human crowds}}.
\newblock {\emph{\JournalTitle{Science Advances}}}
  \textbf{\bibinfo{volume}{7}}, \bibinfo{pages}{eabe7758},
  \url{https://doi.org/10.1126/sciadv.abe7758} (\bibinfo{year}{2021}).

\bibitem{ExpJuelich}
\bibinfo{title}{{Forschungszentrum Jülich, J{\"u}lich Supercomputing Centre.
  Data archive of experiments on pedestrian dynamic}},
  \url{http://ped.fz-juelich.de/dataarchive}.
\newblock \bibinfo{note}{Accessed: 2021-08-01}.

\bibitem{pellegrini2009you}
\bibinfo{author}{Pellegrini, S.}, \bibinfo{author}{Ess, A.},
  \bibinfo{author}{Schindler, K.} \& \bibinfo{author}{Van~Gool, L.}
\newblock \bibinfo{title}{You'll never walk alone: Modeling social behavior for
  multi-target tracking}.
\newblock In \emph{\bibinfo{booktitle}{2009 IEEE 12th international conference
  on computer vision}}, \bibinfo{pages}{261--268},
  \url{https://doi.org/10.1109/ICCV.2009.5459260}
  (\bibinfo{organization}{IEEE}, \bibinfo{year}{2009}).

\bibitem{lerner2007crowds}
\bibinfo{author}{Lerner, A.}, \bibinfo{author}{Chrysanthou, Y.} \&
  \bibinfo{author}{Lischinski, D.}
\newblock \bibinfo{title}{Crowds by example}.
\newblock In \emph{\bibinfo{booktitle}{Computer graphics forum}},
  vol.~\bibinfo{volume}{26}, \bibinfo{pages}{655--664},
  \url{https://doi.org/10.1111/j.1467-8659.2007.01089.x}
  (\bibinfo{organization}{Wiley Online Library}, \bibinfo{year}{2007}).

\bibitem{best2014densesense}
\bibinfo{author}{Best, A.}, \bibinfo{author}{Narang, S.},
  \bibinfo{author}{Curtis, S.} \& \bibinfo{author}{Manocha, D.}
\newblock \bibinfo{title}{Densesense: Interactive crowd simulation using
  density-dependent filters.}
\newblock In \emph{\bibinfo{booktitle}{Symposium on Computer Animation}},
  \bibinfo{pages}{97--102}, \url{https://doi.org/10.2312/sca.20141127}
  (\bibinfo{year}{2014}).

\bibitem{cordes2023dimensionless}
\bibinfo{author}{Cordes, J.}, \bibinfo{author}{Schadschneider, A.} \&
  \bibinfo{author}{Nicolas, A.}
\newblock \bibinfo{journal}{\bibinfo{title}{Dimensionless numbers reveal
  distinct regimes in the structure and dynamics of pedestrian crowds}}.
\newblock {\emph{\JournalTitle{PNAS nexus}}} \textbf{\bibinfo{volume}{3}},
  \bibinfo{pages}{pgae120}, \url{https://doi.org/10.1093/pnasnexus/pgae120}
  (\bibinfo{year}{2024}).

\bibitem{robicquet2016learning}
\bibinfo{author}{Robicquet, A.}, \bibinfo{author}{Sadeghian, A.},
  \bibinfo{author}{Alahi, A.} \& \bibinfo{author}{Savarese, S.}
\newblock \bibinfo{title}{Learning social etiquette: Human trajectory
  understanding in crowded scenes}.
\newblock In \emph{\bibinfo{booktitle}{Computer Vision--ECCV 2016: 14th
  European Conference, Amsterdam, The Netherlands, October 11-14, 2016,
  Proceedings, Part VIII 14}}, \bibinfo{pages}{549--565},
  \url{https://doi.org/10.1007/978-3-319-46484-8_33}
  (\bibinfo{organization}{Springer}, \bibinfo{year}{2016}).

\bibitem{zhou2012understanding}
\bibinfo{author}{Zhou, B.}, \bibinfo{author}{Wang, X.} \&
  \bibinfo{author}{Tang, X.}
\newblock \bibinfo{title}{Understanding collective crowd behaviors: Learning a
  mixture model of dynamic pedestrian-agents}.
\newblock In \emph{\bibinfo{booktitle}{2012 IEEE Conference on Computer Vision
  and Pattern Recognition}}, \bibinfo{pages}{2871--2878},
  \url{https://doi.org/10.1109/CVPR.2012.6248013}
  (\bibinfo{organization}{IEEE}, \bibinfo{year}{2012}).

\bibitem{majecka2009statistical}
\bibinfo{author}{Majecka, B.}
\newblock \bibinfo{journal}{\bibinfo{title}{Statistical models of pedestrian
  behaviour in the forum}}.
\newblock {\emph{\JournalTitle{Master's thesis, School of Informatics,
  University of Edinburgh}}}  (\bibinfo{year}{2009}).
\newblock
  \bibinfo{note}{\href{https://www.semanticscholar.org/paper/Statistical-models-of-pedestrian-behaviour-in-the-Majecka/6505d259758fc2fd4e60da018c35d687a2ddc250}{weblink}}.

\bibitem{wirz2013probing}
\bibinfo{author}{Wirz, M.} \emph{et~al.}
\newblock \bibinfo{journal}{\bibinfo{title}{Probing crowd density through
  smartphones in city-scale mass gatherings}}.
\newblock {\emph{\JournalTitle{EPJ Data Science}}}
  \textbf{\bibinfo{volume}{2}}, \bibinfo{pages}{1--24},
  \url{https://doi.org/10.1140/epjds17} (\bibinfo{year}{2013}).

\bibitem{corbetta2014high}
\bibinfo{author}{Corbetta, A.}, \bibinfo{author}{Bruno, L.},
  \bibinfo{author}{Muntean, A.} \& \bibinfo{author}{Toschi, F.}
\newblock \bibinfo{journal}{\bibinfo{title}{High statistics measurements of
  pedestrian dynamics}}.
\newblock {\emph{\JournalTitle{Transportation Research Procedia}}}
  \textbf{\bibinfo{volume}{2}}, \bibinfo{pages}{96--104},
  \url{https://doi.org/10.1016/j.trpro.2014.09.013} (\bibinfo{year}{2014}).

\bibitem{corbetta2020high}
\bibinfo{author}{Corbetta, A.}, \bibinfo{author}{Schilders, L.} \&
  \bibinfo{author}{Toschi, F.}
\newblock \bibinfo{journal}{\bibinfo{title}{High-statistics modeling of complex
  pedestrian avoidance scenarios}}.
\newblock {\emph{\JournalTitle{Crowd Dynamics, Volume 2: Theory, Models, and
  Applications}}} \bibinfo{pages}{33--53},
  \url{https://doi.org/10.1007/978-3-030-50450-2_3} (\bibinfo{year}{2020}).

\bibitem{amirian2020opentraj}
\bibinfo{author}{Amirian, J.} \emph{et~al.}
\newblock \bibinfo{title}{Opentraj: Assessing prediction complexity in human
  trajectories datasets}.
\newblock In \emph{\bibinfo{booktitle}{Asian Conference on Computer Vision
  (ACCV)}}, \bibinfo{number}{CONF},
  \url{https://doi.org/10.1007/978-3-030-69544-6_34}
  (\bibinfo{organization}{Springer}, \bibinfo{year}{2020}).

\bibitem{haghani2020empiricalII}
\bibinfo{author}{Haghani, M.}
\newblock \bibinfo{journal}{\bibinfo{title}{Empirical methods in pedestrian,
  crowd and evacuation dynamics: Part ii. field methods and controversial
  topics}}.
\newblock {\emph{\JournalTitle{Safety science}}}
  \textbf{\bibinfo{volume}{129}}, \bibinfo{pages}{104760},
  \url{https://doi.org/10.1016/j.ssci.2020.104760} (\bibinfo{year}{2020}).

\bibitem{korbmacher2022}
\bibinfo{author}{Korbmacher, R.} \& \bibinfo{author}{Tordeux, A.}
\newblock \bibinfo{journal}{\bibinfo{title}{Review of pedestrian trajectory
  prediction methods: Comparing deep learning and knowledge-based approaches}}.
\newblock {\emph{\JournalTitle{IEEE Transactions on Intelligent Transportation
  Systems}}} \textbf{\bibinfo{volume}{23}}, \bibinfo{pages}{24126--24144},
  \url{https://doi.org/10.1109/TITS.2022.3205676} (\bibinfo{year}{2022}).

\bibitem{LeProgres2022}
\bibinfo{author}{{Le Progr\`es}}.
\newblock \bibinfo{journal}{\bibinfo{title}{F\^ete des lumi\`eres 2022: Plus de
  2 millions de visiteurs cette ann\'ee}}.
\newblock {\emph{\JournalTitle{Le Progr\`es}}}  (\bibinfo{year}{2022}).
\newblock
  \bibinfo{note}{\href{https://www.leprogres.fr/culture-loisirs/2022/12/12/plus-de-2-millions-de-visiteurs-a-la-fete-des-lumieres-la-mairie-affiche-de-bons-chiffres-de-frequentation}{weblink}}.

\bibitem{ActuLyon2023}
\bibinfo{author}{Zuili, T.}
\newblock \bibinfo{journal}{\bibinfo{title}{F\^ete des lumi\`eres: Colis
  suspects, foule, pannes... comment se pr\'epare tcl à lyon}}.
\newblock {\emph{\JournalTitle{ActuLyon}}}  (\bibinfo{year}{2023}).
\newblock
  \bibinfo{note}{\href{https://actu.fr/auvergne-rhone-alpes/lyon_69123/fete-des-lumieres-colis-suspects-foule-pannes-comment-se-prepare-tcl-a-lyon_60418009.html}{weblink}}.

\bibitem{liang2024unraveling}
\bibinfo{author}{Liang, H.}, \bibinfo{author}{Lee, S.}, \bibinfo{author}{Sun,
  J.} \& \bibinfo{author}{Wong, S.}
\newblock \bibinfo{journal}{\bibinfo{title}{Unraveling the causes of the seoul
  halloween crowd-crush disaster}}.
\newblock {\emph{\JournalTitle{PLoS one}}} \textbf{\bibinfo{volume}{19}},
  \bibinfo{pages}{e0306764}, \url{https://doi.org/10.1371/journal.pone.0306764}
  (\bibinfo{year}{2024}).

\bibitem{data_madras_Geometry}
\bibinfo{author}{Dufour, O.} \emph{et~al.}
\newblock \bibinfo{title}{{Dense Crowd Dynamics and Pedestrian Trajectories: A
  Multiscale Field Study at the Fête des Lumières in Lyon >> Geometry.csv}},
  \url{10.5281/zenodo.13830435},
  \urlprefix\url{https://doi.org/10.5281/zenodo.13830435}
  (\bibinfo{year}{2024}).

\bibitem{data_madras_GPS}
\bibinfo{author}{Dufour, O.} \emph{et~al.}
\newblock \bibinfo{title}{{Dense Crowd Dynamics and Pedestrian Trajectories: A
  Multiscale Field Study at the Fête des Lumières in Lyon >> GPS traces and
  physical contacts }}, \url{10.5281/zenodo.13830435},
  \urlprefix\url{https://doi.org/10.5281/zenodo.13830435}
  (\bibinfo{year}{2024}).

\bibitem{data_madras_LargeViewTrajectories}
\bibinfo{author}{Dufour, O.} \emph{et~al.}
\newblock \bibinfo{title}{{Dense Crowd Dynamics and Pedestrian Trajectories: A
  Multiscale Field Study at the Fête des Lumières in Lyon >> LargeView
  trajectories}}, \url{10.5281/zenodo.13830435},
  \urlprefix\url{https://doi.org/10.5281/zenodo.13830435}
  (\bibinfo{year}{2024}).

\bibitem{data_madras_surveys}
\bibinfo{author}{Dufour, O.} \emph{et~al.}
\newblock \bibinfo{title}{{Dense Crowd Dynamics and Pedestrian Trajectories: A
  Multiscale Field Study at the Fête des Lumières in Lyon >> Surveys }},
  \url{10.5281/zenodo.13830435},
  \urlprefix\url{https://doi.org/10.5281/zenodo.13830435}
  (\bibinfo{year}{2024}).

\bibitem{data_madras_TopViewTrajectories}
\bibinfo{author}{Dufour, O.} \emph{et~al.}
\newblock \bibinfo{title}{{Dense Crowd Dynamics and Pedestrian Trajectories: A
  Multiscale Field Study at the Fête des Lumières in Lyon >> TopView
  trajectories}}, \url{10.5281/zenodo.13830435},
  \urlprefix\url{https://doi.org/10.5281/zenodo.13830435}
  (\bibinfo{year}{2024}).

\bibitem{boltesCollectingPedestrianTrajectories2013}
\bibinfo{author}{Boltes, M.} \& \bibinfo{author}{Seyfried, A.}
\newblock \bibinfo{journal}{\bibinfo{title}{Collecting pedestrian
  trajectories}}.
\newblock {\emph{\JournalTitle{Neurocomputing}}}
  \textbf{\bibinfo{volume}{100}}, \bibinfo{pages}{127--133},
  \url{https://doi.org/10.1016/j.neucom.2012.01.036} (\bibinfo{year}{2013}).

\bibitem{boltes_PeTrack_2022}
\bibinfo{author}{Boltes, M.} \emph{et~al.}
\newblock \bibinfo{title}{{{PeTrack}}}.
\newblock \bibinfo{howpublished}{Zenodo},
  \url{https://doi.org/10.5281/zenodo.6320753} (\bibinfo{year}{2022}).

\bibitem{boltes2010automatic}
\bibinfo{author}{Boltes, M.}, \bibinfo{author}{Seyfried, A.},
  \bibinfo{author}{Steffen, B.} \& \bibinfo{author}{Schadschneider, A.}
\newblock \bibinfo{title}{Automatic extraction of pedestrian trajectories from
  video recordings}.
\newblock In \emph{\bibinfo{booktitle}{Pedestrian and evacuation dynamics
  2008}}, \bibinfo{pages}{43--54},
  \url{https://doi.org/10.1007/978-3-642-04504-2_3}
  (\bibinfo{organization}{Springer}, \bibinfo{year}{2010}).

\bibitem{johansson2008crowd}
\bibinfo{author}{Johansson, A.}, \bibinfo{author}{Helbing, D.},
  \bibinfo{author}{Al-Abideen, H.~Z.} \& \bibinfo{author}{Al-Bosta, S.}
\newblock \bibinfo{journal}{\bibinfo{title}{From crowd dynamics to crowd
  safety: a video-based analysis}}.
\newblock {\emph{\JournalTitle{Advances in Complex Systems}}}
  \textbf{\bibinfo{volume}{11}}, \bibinfo{pages}{497--527},
  \url{https://doi.org/10.1142/S0219525908001854} (\bibinfo{year}{2008}).

\bibitem{streamlit}
\bibinfo{author}{Mohcine, C.} \& \bibinfo{author}{Dufour, O.}
\newblock \bibinfo{title}{{MADRAS-data-app}}, \url{10.5281/zenodo.10694867}, \urlprefix\url{https://go.fzj.de/madras-app}
  (\bibinfo{year}{2024}).

\bibitem{pedpy}
\bibinfo{author}{{Schrödter, Tobias}} \& \bibinfo{author}{{The PedPy
  Development Team}}.
\newblock \bibinfo{title}{Pedestriandynamics/pedpy: v1.0.2},
  \url{https://doi.org/10.5281/zenodo.10016938} (\bibinfo{year}{2023}).

\bibitem{OfficialMap}
\bibinfo{title}{{Fête des Lumière 2022 Official Map}}.
\newblock
  \bibinfo{note}{\href{https://www.fetedeslumieres.lyon.fr/en/map/2022-map}{weblink}}.

\bibitem{van2011reciprocal}
\bibinfo{author}{Van Den~Berg, J.}, \bibinfo{author}{Guy, S.~J.},
  \bibinfo{author}{Lin, M.} \& \bibinfo{author}{Manocha, D.}
\newblock \bibinfo{title}{Reciprocal n-body collision avoidance}.
\newblock In \emph{\bibinfo{booktitle}{Robotics Research: The 14th
  International Symposium ISRR}}, \bibinfo{pages}{3--19},
  \url{https://doi.org/10.1007/978-3-642-19457-3_1}
  (\bibinfo{organization}{Springer}, \bibinfo{year}{2011}).

\bibitem{karamouzas2017implicit}
\bibinfo{author}{Karamouzas, I.}, \bibinfo{author}{Sohre, N.},
  \bibinfo{author}{Narain, R.} \& \bibinfo{author}{Guy, S.~J.}
\newblock \bibinfo{journal}{\bibinfo{title}{Implicit crowds: Optimization
  integrator for robust crowd simulation}}.
\newblock {\emph{\JournalTitle{ACM Transactions on Graphics (TOG)}}}
  \textbf{\bibinfo{volume}{36}}, \bibinfo{pages}{1--13},
  \url{https://doi.org/10.1145/3072959.3073705} (\bibinfo{year}{2017}).

\bibitem{vanumu2017fundamental}
\bibinfo{author}{Vanumu, L.~D.}, \bibinfo{author}{Ramachandra~Rao, K.} \&
  \bibinfo{author}{Tiwari, G.}
\newblock \bibinfo{journal}{\bibinfo{title}{Fundamental diagrams of pedestrian
  flow characteristics: A review}}.
\newblock {\emph{\JournalTitle{European transport research review}}}
  \textbf{\bibinfo{volume}{9}}, \bibinfo{pages}{1--13},
  \url{https://doi.org/10.1007/s12544-017-0264-6} (\bibinfo{year}{2017}).

\bibitem{jiang2014traffic}
\bibinfo{author}{Jiang, R.} \emph{et~al.}
\newblock \bibinfo{journal}{\bibinfo{title}{Traffic experiment reveals the
  nature of car-following}}.
\newblock {\emph{\JournalTitle{PloS one}}} \textbf{\bibinfo{volume}{9}},
  \bibinfo{pages}{e94351}, \url{https://doi.org/10.1371/journal.pone.0094351}
  (\bibinfo{year}{2014}).

\bibitem{nicolas2023social}
\bibinfo{author}{Nicolas, A.} \& \bibinfo{author}{Hassan, F.~H.}
\newblock \bibinfo{journal}{\bibinfo{title}{Social groups in pedestrian crowds:
  review of their influence on the dynamics and their modelling}}.
\newblock {\emph{\JournalTitle{Transportmetrica A: transport science}}}
  \textbf{\bibinfo{volume}{19}}, \bibinfo{pages}{1970651},
  \url{https://doi.org/10.1080/23249935.2021.1970651} (\bibinfo{year}{2023}).

\bibitem{wang2023exploring}
\bibinfo{author}{Wang, J.}, \bibinfo{author}{Lv, W.}, \bibinfo{author}{Jiang,
  H.}, \bibinfo{author}{Fang, Z.} \& \bibinfo{author}{Ma, J.}
\newblock \bibinfo{journal}{\bibinfo{title}{Exploring crowd persistent dynamism
  from pedestrian crossing perspective: An empirical study}}.
\newblock {\emph{\JournalTitle{Transportation research part C: emerging
  technologies}}} \textbf{\bibinfo{volume}{157}}, \bibinfo{pages}{104400},
  \url{https://doi.org/10.1016/j.trc.2023.104400} (\bibinfo{year}{2023}).

\bibitem{helbing1997modelling}
\bibinfo{author}{Helbing, D.}, \bibinfo{author}{Keltsch, J.} \&
  \bibinfo{author}{Molnar, P.}
\newblock \bibinfo{journal}{\bibinfo{title}{Modelling the evolution of human
  trail systems}}.
\newblock {\emph{\JournalTitle{Nature}}} \textbf{\bibinfo{volume}{388}},
  \bibinfo{pages}{47--50}, \url{https://doi.org/10.1038/40353}
  (\bibinfo{year}{1997}).

\bibitem{echeverria2023body}
\bibinfo{author}{Echeverr{\'\i}a-Huarte, I.} \& \bibinfo{author}{Nicolas, A.}
\newblock \bibinfo{journal}{\bibinfo{title}{Body and mind: Decoding the
  dynamics of pedestrians and the effect of smartphone distraction by coupling
  mechanical and decisional processes}}.
\newblock {\emph{\JournalTitle{Transportation research part C: emerging
  technologies}}} \textbf{\bibinfo{volume}{157}}, \bibinfo{pages}{104365},
  \url{https://doi.org/10.1016/j.trc.2023.104365} (\bibinfo{year}{2023}).

\bibitem{dey2012spatial}
\bibinfo{author}{Dey, S.}, \bibinfo{author}{Das, D.} \&
  \bibinfo{author}{Rajesh, R.}
\newblock \bibinfo{journal}{\bibinfo{title}{Spatial structures and giant number
  fluctuations in models of active matter}}.
\newblock {\emph{\JournalTitle{Physical review letters}}}
  \textbf{\bibinfo{volume}{108}}, \bibinfo{pages}{238001},
  \url{https://doi.org/10.1103/PhysRevLett.108.238001} (\bibinfo{year}{2012}).

\bibitem{manning2023essay}
\bibinfo{author}{Manning, M.~L.}
\newblock \bibinfo{journal}{\bibinfo{title}{Essay: Collections of deformable
  particles present exciting challenges for soft matter and biological
  physics}}.
\newblock {\emph{\JournalTitle{Physical Review Letters}}}
  \textbf{\bibinfo{volume}{130}}, \bibinfo{pages}{130002},
  \url{https://doi.org/10.1103/PhysRevLett.130.130002} (\bibinfo{year}{2023}).

\bibitem{nicolas2019mechanical}
\bibinfo{author}{Nicolas, A.}, \bibinfo{author}{Kuperman, M.},
  \bibinfo{author}{Iba{\~n}ez, S.}, \bibinfo{author}{Bouzat, S.} \&
  \bibinfo{author}{Appert-Rolland, C.}
\newblock \bibinfo{journal}{\bibinfo{title}{Mechanical response of dense
  pedestrian crowds to the crossing of intruders}}.
\newblock {\emph{\JournalTitle{Scientific reports}}}
  \textbf{\bibinfo{volume}{9}}, \bibinfo{pages}{105},
  \url{https://doi.org/10.1038/s41598-018-36711-7} (\bibinfo{year}{2019}).

\end{thebibliography}
\end{document}